\documentclass[twocolumn,prl,amsmath,amssymb, aps,superscriptaddress, nofootinbib]{revtex4-2}
\usepackage{comment}
\usepackage[titletoc,toc,title]{appendix}
\usepackage{titletoc}
\usepackage{titlesec}
\usepackage[version=4]{mhchem} 
\usepackage[normalem]{ulem}
\usepackage{mhchem}
\usepackage{amsmath,bm}
\usepackage{amssymb}
\usepackage{amsfonts}
\usepackage{afterpage}
\usepackage{graphicx}
\usepackage{xfrac}
\usepackage{footmisc}
\usepackage{comment}
\usepackage{pifont}
\usepackage{times}
\usepackage[hidelinks]{hyperref}
\usepackage{enumitem}
\usepackage{centernot}
\usepackage{cancel}
\usepackage{slashed}
\usepackage{relsize}
\usepackage{amsmath,amssymb,mathrsfs}
\usepackage{times}
\usepackage{epsfig}
\usepackage{verbatim}
\usepackage{bm}
\usepackage[utf8]{inputenc}
\usepackage{graphics}
\usepackage{graphicx,epsfig,amssymb,amsmath,color,cancel}
\usepackage{subfigure}
\usepackage[english]{babel}
\usepackage{adjustbox}
\usepackage{physics}
\usepackage{placeins}

\usepackage{array}
\newcolumntype{P}[1]{>{\centering\arraybackslash}p{#1}}
\newcolumntype{M}[1]{>{\centering\arraybackslash}m{#1}}

\usepackage[table]{xcolor}
\usepackage{tabularray}
\usepackage{fontawesome}

\UseTblrLibrary{booktabs}

\definecolor{zima_blue}{HTML}{1393C1}
\hypersetup{setpagesize=false,bookmarksnumbered=true,bookmarksopen=true,%
colorlinks=true,linkcolor=zima_blue,urlcolor=zima_blue,citecolor=zima_blue,linktocpage=false}

\newcommand{\vphi}{v_{\phi}}

\usepackage{tikz}

\usepackage{orcidlink}

\usepackage[starfontserif 
]{starfont}
\usepackage{amsmath}

\DeclareSymbolFont{starfontsym}{OT1}{sts}{m}{n}
\DeclareMathSymbol{\mathSun}{\mathord}{starfontsym}{115}
\DeclareMathSymbol{\mathMercury}{\mathord}{starfontsym}{102}
\DeclareMathSymbol{\mathVenus}{\mathord}{starfontsym}{103}
\DeclareMathSymbol{\mathTerra}{\mathord}{starfontsym}{76}
\DeclareMathSymbol{\mathvarTerra}{\mathord}{starfontsym}{108}
\DeclareMathSymbol{\mathMoon}{\mathord}{starfontsym}{100}
\DeclareMathSymbol{\mathvarMoon}{\mathord}{starfontsym}{97}
\DeclareMathSymbol{\mathMars}{\mathord}{starfontsym}{104}
\DeclareMathSymbol{\mathJupiter}{\mathord}{starfontsym}{106}
\DeclareMathSymbol{\mathSaturn}{\mathord}{starfontsym}{83}
\DeclareMathSymbol{\mathUranus}{\mathord}{starfontsym}{70}
\DeclareMathSymbol{\mathvarUranus}{\mathord}{starfontsym}{65}
\DeclareMathSymbol{\mathNeptune}{\mathord}{starfontsym}{71}
\DeclareMathSymbol{\mathPluto}{\mathord}{starfontsym}{74}
\DeclareMathSymbol{\mathvarPluto}{\mathord}{starfontsym}{72}

\titleformat{\section}
{\normalfont\fontsize{10}{12}\bfseries  \centering }{\thesection.}{1em}{}
\titleformat{\subsection}
{\normalfont\fontsize{10}{12}\bfseries \centering}{\thesubsection.}{1em}{}
\titleformat{\subsubsection}
{\normalfont\fontsize{10}{12}\bfseries \centering}{\thesubsubsection)}{1em}{}

\titleformat{\paragraph}
{\normalfont\fontsize{10}{12}\bfseries  }{\thesection:}{1em}{}

\definecolor{linkcolor}{rgb}{0.7752941176470588, 0.22078431372549023, 0.2262745098039215}
\definecolor{linkcolor}{HTML}{1393C1}
\newcommand{\githubmaster}[3][linkcolor]{\href{#2}{\color{#1}{#3}}}%

\begin{document}


\title{Primordial Black Holes from Conformal Higgs}

\author{Yann Gouttenoire~\orcidlink{0000-0003-2225-6704}}
\email{yann.gouttenoire@gmail.com}
\affiliation{School of Physics and Astronomy, Tel-Aviv University, Tel-Aviv 69978, Israel}

\begin{abstract}

Scale-invariant extensions of the electroweak theory are not only attractive because they can dynamically generate the weak scale, but also due to their role in facilitating supercooled first-order phase transitions.
We study the minimal scale-invariant $U(1)_{\rm D}$ extension of the standard model and show that Primordial Black Holes (PBHs) can be abundantly produced. The mass of these PBHs is bounded from above by that of the moon due to QCD catalysis limiting the amount of supercooling. Lunar-mass PBHs, which are produced for dark Higgs vev $v_\phi\simeq 20~\rm TeV$, correspond to the best likelihood to explain the HSC lensing anomaly. For $v_\phi\gtrsim 400~\rm TeV$, the model can explain hundred per cent of dark matter. At even larger hierarchy of scales, it can contribute to the $511~\rm keV$ line.
While the gravitational wave (GW) signal produced by the HSC anomaly interpretation is large and detectable by LISA above astrophysical foreground, the dark matter interpretation in terms of PBHs can not be entirely probed by future GW detection. This is due to the dilution of the signal by the entropy injected during the decay of the long-lived $U(1)_{\rm D}$ scalar. This extended lifetime is a natural consequence of the large hierarchy of scales.

\vspace*{5pt} \noindent \textbf{\texttt{GitHub}}: The \textsc{CCBounce} code to calculate bounce action and tunneling temperature in single-field cosmological first-order phase transition models is available at \href{https://github.com/YannGou/CCBounce}{\faGithub\;YannGou/CCBounce}.

\vspace*{5pt}
\noindent
DOI:~\href{https://doi.org/10.1016/j.physletb.2024.138800}{ 10.1016/j.physletb.2024.138800}

\end{abstract}

\maketitle


\section{INTRODUCTION}
Stellar-mass and supermassive black holes are at the forefront of astrophysical research, particularly following the detection of  gravitational waves by LIGO/Virgo \cite{LIGOScientific:2021djp} and Pulsar Timing Arrays (PTA) \cite{NANOGrav:2023gor,EPTA:2023fyk,Reardon:2023gzh}. However, general relativity allows black holes to have any
mass. Sub-solar mass black holes can form from the gravitational collapse of large overdensities in the primordial plasma~\cite{Carr:1974nx} and could explain $100~\%$ of the observed dark matter relic density in the mass range $10^{-16}M_{\odot}\lesssim M_{\rm PBH}\lesssim 10^{-10}M_{\odot}$ \cite{Carr:2016drx,Carr:2020xqk,Carr:2020gox,Green:2020jor}. The production of  Primordial Black Holes (PBHs) by ``late-blooming'' during supercooled first-order phase transition, introduced in 1981 \cite{Sato:1981bf,Maeda:1981gw,Sato:1981gv,Kodama:1981gu,Kodama:1982sf,Hsu:1990fg}, has received considerable attention recently \cite{Liu:2021svg,Hashino:2021qoq,He:2022amv,Kawana:2022olo,Lewicki:2023ioy,Gouttenoire:2023naa,Baldes:2023rqv,Salvio:2023ynn,Banerjee:2023qya,Gehrman:2023esa,Flores:2024lng,Lewicki:2024ghw}, also in relation with the PTA signal \cite{NANOGrav:2023hvm,Gouttenoire:2023bqy,He:2023ado,Ellis:2023oxs}.

In this \textit{letter}, we investigate whether the electroweak phase transition (EWPT), minimally extended by new physics, can generate PBHs.
Apart from its second derivative around $\left<H\right>\simeq 246~$GeV measured in 2011 \cite{ATLAS:2012yve,CMS:2012qbp}, the shape of the Higgs potential remains unconstrained \cite{Salam:2022izo}. In 1976, Gildener and S. Weinberg \cite{Gildener:1976ih} suggested that the Higgs mass could vanish at tree level and be dynamically generated when quantum loop corrections, introduced by Coleman and E. Weinberg (CW) in 1973 \cite{Coleman:1973jx}, drive the Higgs quartic to negative values in the infrared. The original model has been ruled out by the large top mass \cite{Hempfling:1996ht}, necessitating the introduction of additional bosonic fields into the Standard Model (SM) to make it viable. One of the simplest scale-invariant extensions of electroweak theory involves a complex scalar field $\phi$, neutral under the SM but charged under an additional $U(1)_{\rm D}$ gauge group \cite{Hempfling:1996ht}. This model has attracted considerable attention due to its small number of new parameters and potential implications for dark matter, leptogenesis and neutrino oscillations \cite{Chang:2007ki,Foot:2007as,Alexander-Nunneley:2010tyr,Englert:2013gz,Khoze:2013uia,Khoze:2014xha,Lindner:2014oea,Humbert:2015epa,Oda:2015gna,Das:2016zue}. Phase transitions driven by a Coleman-Weinberg potential plus thermal corrections:
\begin{equation}
\label{eq:CW_potential_intro}
V(\phi) \simeq \frac{g^2T^2\phi^2}{2} +  \frac{ \beta_\lambda\phi^4}{4} \left[\log{\left( \frac{\phi}{\vphi}\right) - \frac{1}{4}}\right],
\end{equation} 
are known to be strongly first order (1stOPT) and supercooled \cite{Guth:1980zk,Witten:1980ez}. We have Taylor-expanded thermal corrections around the symmetric minima $\phi=0$. We have introduced the beta function $\beta_\lambda$ of the quartic coupling $\lambda \phi^4$ and the coupling $g$ with particles of the plasma. For the $U(1)_{\rm D}$ model with gauge coupling $g_{\rm D}$ studied in this \textit{letter}, one has $\beta_\lambda\simeq 3g_{\rm D}^4/8\pi^2$ and $g^2\simeq g_{\rm D}^2/12$. Spontaneous breaking $\left<\phi\right>\neq 0$ is communicated to the Higgs field $H$ through the mixing $\lambda |H|^2\phi^2$. The scalar $\phi$ can be interpreted as the ``Higgs of the Higgs'' \cite{Hambye:2013}.  Thanks to the logarithmic potential, the temperature $T_n$ at which bubbles nucleate and percolate is exponentially suppressed for $\beta_\lambda \ll 1$:
\begin{equation}
\label{eq:Teq_def}
T_n~ \simeq~T_c\, \exp\left(-\frac{16\pi}{3 S_{\rm c}} \frac{g}{\beta_\lambda} \right),\quad S_c \simeq 130,
\end{equation}
with respect to the critical temperature $T_c$ when the 1stOPT becomes energetically allowed. $S_c$ is the bounce action at percolation. Coleman-Weinberg dominance in the potential in Eq.~\eqref{eq:CW_potential_intro} additionally implies that the bounce action cost $S_3/T$ -- which controls the bubble nucleation rate per unit of volume $\Gamma_{\mathsmaller{\rm V}} = T^4 \exp(S_3/T)$ -- decreases only logarithmically with the temperature. This implies that the completion rate of the PT:
\begin{equation}
\label{eq:beta_def}
\beta ~\equiv~\frac{d\log{\Gamma_{\mathsmaller{\rm V}}}}{dt} \simeq \left(\frac{3\beta_{\lambda}}{16\pi g}S_c^2- 4\right)H,
\end{equation}
can be relatively small for $\beta_\lambda \ll 1$.

 We investigate the minimal $U(1)_D$ scale-invariant extension of the electroweak theory. We study the tunneling dynamics both numerically \cite{cppbounce} and analytically taking into account QCD catalysis and collider constraints. We find that the two necessary conditions for the formation of PBHs in observable quantities {derived in the companion paper \cite{Gouttenoire:2023naa}} -- significant supercooling ($T_n \ll T_c$) and a slow rate of completion ($\beta/H \lesssim 7$) -- can be satisfied as soon as the Dark Higgs Vacuum Expectation Value (VEV) is larger than $v_{\phi} \gtrsim 10~\rm TeV$. 
The gravitational wave detection prospects are calculated in all the parameter space, accounting for astrophysical foregrounds. We account for the matter era following the phase transition resulting from the dark Higgs becoming long-lived at large VEV.


\section{SCALE INVARIANT EXTENSION OF THE ELECTROWEAK THEORY}
\underline{The scalar potential} ---
We extend the SM with a complex scalar field $\Phi$ charged under a hidden $U(1)_{\rm D}$ dark photon $V_\mu$ with coupling constant $g_{\rm D}$   \cite{Hempfling:1996ht}:
\begin{equation}
\mathcal{L}_{\rm tree} =  \left|D_\mu \Phi\right|^2   -V_{\rm tree}(|\Phi|,|H|)  ,
\end{equation}
with $D_\mu = \partial_\mu + ig_{\rm D} t^a V_\mu^a$. Supposing Bardeen's conjecture that scale-invariance is a fundamental symmetry of Nature at tree level \cite{Bardeen:1995kv}, the only allowed operators in the scalar potential are:
\begin{equation}
V_{\rm tree}(|\Phi|,|H|) =   \lambda_{h}|H|^{4}  +\lambda_{\phi} |\Phi|^{4} -\lambda_{ h\phi} |\Phi|^{2} |H|^{2},
\end{equation}
where $\lambda_{ h \phi}$ is the mixing coupling with the SM Higgs $H$. In the unitarity gauge, we can parameterize the scalar fields as:
\begin{equation}
H =  \left(0,\frac{h}{\sqrt{2}}\right), \qquad \Phi =  \frac{\phi }{\sqrt{2}}.
\end{equation} 
Scale-invariance is violated by quantum interactions encoded in Renormalized Group Equations (RGE), see e.g. \cite{Khoze:2014xha}.
In App.~\ref{app:nearly_conf_sec}, we show that upon assuming the ordering $g_{\rm D}^2 \gg \lambda_\phi, \lambda_{h\phi}$, the 1-loop corrected potential is well approximated by its leading-log term:
\begin{equation}
\label{eq:CW_potential}
V_{T=0}(\phi)  = \beta_\lambda \frac{\phi^4}{4}\left[\textrm{log}\left( \frac{\phi}{\vphi}\right) - \frac{1}{4}\right], \quad \beta_\lambda\simeq 6 \alpha_{\rm D}^2.
\end{equation}
We use the full RGE improved coupling $\lambda_\phi(\phi)$ in our numerical computations. The Coleman-Weinberg potential in Eq.~\eqref{eq:CW_potential} generates a VEV for the $U(1)_{\rm D}$ field $\left<\phi \right> = \vphi$, which through the mixing $\lambda_{ h\phi}h^2\phi^2$ generates a VEV for the Higgs $\left<h\right> = v_{\rm EW}$. Fixing the Higgs VEV to $v_{\rm EW} =246~\rm GeV$ and its mass to $m_h = 125~\rm GeV$ implies:
\begin{equation}
\label{eq:Higgs_mixing}
\lambda_h=\frac{m_h^2}{2v_{\rm EW}^2}\simeq 0.13,\quad \lambda_{h\phi} =\frac{m_h^2}{v_{\phi}^2} \simeq 0.02 \left( \frac{1~\rm TeV}{\vphi}\right)^2.
\end{equation} 
At finite temperature the potential for $\phi$ receives thermal corrections which under the ordering $g_{\rm D}^2 \gg \lambda_\phi, \lambda_{h\phi}$ are dominated by the dark photon $V_\mu$.  The 1-loop and Daisy contributions read \cite{Dolan:1973qd}:
\begin{equation}
\label{eq:ThermalPotential}
 V_{\rm T}(\phi) =  \frac{ T^{4}}{2\pi^2} J_{B}\left(\frac{m_{V}^2(\phi)}{T^2}\right)+\frac{T}{12\pi} \left[ m_{V}^3 - \left( m_{V}^{2} + \Pi_{V}\right)^{3/2} \right] ,
\end{equation}
with $m_{V}^2(\phi)=g_{\rm D}^2 \phi^2$ and $\Pi_{V} =  g_{\rm D}^2 T^{2}/6$ \cite{Carrington:1991hz}.\\
\underline{QCD catalysis} ---
In App.~\ref{app:nearly_conf_sec}, we show that the motion of the Higgs field can be safely neglected during the tunneling if $v_{\phi}\gtrsim ~\rm TeV$. An exception arises if the universe super-cools below the temperature of QCD confinement $T_{n} \lesssim \Lambda_{\rm QCD}$. Then the top chiral condensate induces a negative linear slope in the Higgs direction $-\frac{y_t}{\sqrt{2}} \left<t_{\rm L}t_{\rm R}\right> h $ \cite{Witten:1980ez}, see also \cite{Iso:2017uuu,vonHarling:2017yew,Hambye:2018qjv,Baratella:2018pxi,Fujikura:2019oyi,Bodeker:2021mcj,Sagunski:2023ynd}.
The Higgs acquires a VEV:
\begin{equation}
\label{eq:h_QCD_formula}
\left<h\right>_{\mathsmaller{\rm QCD}}^3 = \frac{4y_t}{\sqrt{2}\lambda_h} \left<t_{\rm L}t_{\rm R}\right> \sim \Lambda_{\rm QCD}^3,
\end{equation} 
which through the mixing $\lambda_{h\phi}$ induces a negative mass for the hidden scalar $\phi$:
\begin{align}
\label{eq:QCD_contribution}
V_{\rm QCD}(\phi) = -\frac{1}{4}\lambda_{h\phi}\left<h\right>_{\mathsmaller{\rm QCD}}^2\phi^2{\Theta(\Lambda_{\rm QCD}-T).}
\end{align}
Since both the Higgs quartic $\lambda_h$ and the top Yukawa $y_t$ diverges below $\Lambda_{\rm QCD}$, it is not possible to predict the precise value of $\left<h\right>_{\mathsmaller{\rm QCD}}$ \cite{Hambye:2018qjv,Bodeker:2021mcj}. Waiting for further studies, we approximate the temperature dependence of QCD effects with a $\Theta$ function and fix  $\left<h\right>_{\mathsmaller{\rm QCD}} = 100~\rm MeV$.\\
\underline{Bubble nucleation} --- 
The thermal corrections in Eq.~\eqref{eq:ThermalPotential} generates a barrier between the minima in $\phi=0$ and $\phi=v_\phi$. Thermal fluctuation can drive the field over the barrier with a rate per unit of volume \cite{Linde:1981zj}:
\begin{equation}
\Gamma_{\mathsmaller{\rm V}}(T)\simeq  T^4\left( \frac{S_3/T}{2\pi} \right)^{3/2} {\rm exp} \left(-S_3/T\right) ,
\label{eq:tunneling_rate}
\end{equation}
where $S_3$ is the $O_3$-symmetric bounce action.  
We wrote a C++ undershooting-overshooting algorithm {$\tt CCBounce$} \cite{cppbounce} to find the tunneling trajectory and compute the bounce action $S_3$. We found it to be well approximated by the thick-wall formula:
\begin{align}
\label{eq:S3overT_cw}
&\frac{S_3}{T} \simeq  \frac{A}{\textrm{log}\left(\frac{M}{T}\right)} \quad \text{with} \quad M = 0.43\,g_{\rm D}\,\vphi, \quad \\
&A= \frac{127.3}{g_{\rm D}^3}\sqrt{1 - \left(\frac{T_{\mathsmaller{\rm QCD}}}{T}\right)^{\!2}}, \quad T_{\mathsmaller{\rm QCD}} =  \frac{\sqrt{6\lambda_{\rm h\sigma}}}{g_{\rm D}}\left<h\right>_{\mathsmaller{\rm QCD}}. \notag
\end{align}
$T_{\mathsmaller{\rm QCD}}$ is the temperature when the QCD contribution in Eq.~\eqref{eq:QCD_contribution} becomes larger than the thermal barrier of $\phi$ and below which the EW-PT must necessarily complete. We refer to App.~\ref{app:nearly_conf_sec} for more details on the bounce action calculation and for a comparison of the thick-wall formula with the proper numerical treatment.

Nucleation happens when the decay rate in Eq.~\eqref{eq:tunneling_rate} multiplied by the causal volume becomes comparable to the Hubble expansion rate: $\Gamma_{\mathsmaller{\rm V}}(T_{n}) \simeq H^4(T_{n})$. Plugging Eq.~\eqref{eq:S3overT_cw}, we deduce the nucleation temperature:
\begin{equation}
T_{n}/T_{\rm eq} \simeq 5\,\exp\left[- \frac{127}{g_{\rm D}^3S_c} \right],\quad S_c \simeq 4 \ln{\frac{T_{n}}{H(T_{n})}} \simeq 130,
\end{equation}
where we neglected the QCD contribution for clarity. We use a proper root solving algorithm to find $T_n$ in our plots. In the absence of QCD, effects, the bounce action $S_3/T$ flattens at low temperature, causing the absence of solution for $\Gamma_{\mathsmaller{\rm V}}(T_{n}) \simeq H^4(T_{n})$ below the temperature \cite{DelleRose:2019pgi,Baldes:2021aph,Gouttenoire:2022gwi}:
\begin{equation}
    T_{n}^{\rm min} \simeq \sqrt{M H_n},
\end{equation}
where $H_n\equiv H(T_n)$ and $M$ given by Eq.~\eqref{eq:S3overT_cw}. Instead, QCD confinement generates a sharp decrease of $S_3/T$ down to zero below the temperature $T_{\mathsmaller{\rm QCD}}$.

\begin{figure}[h!]
\centering
\includegraphics[width=240pt]{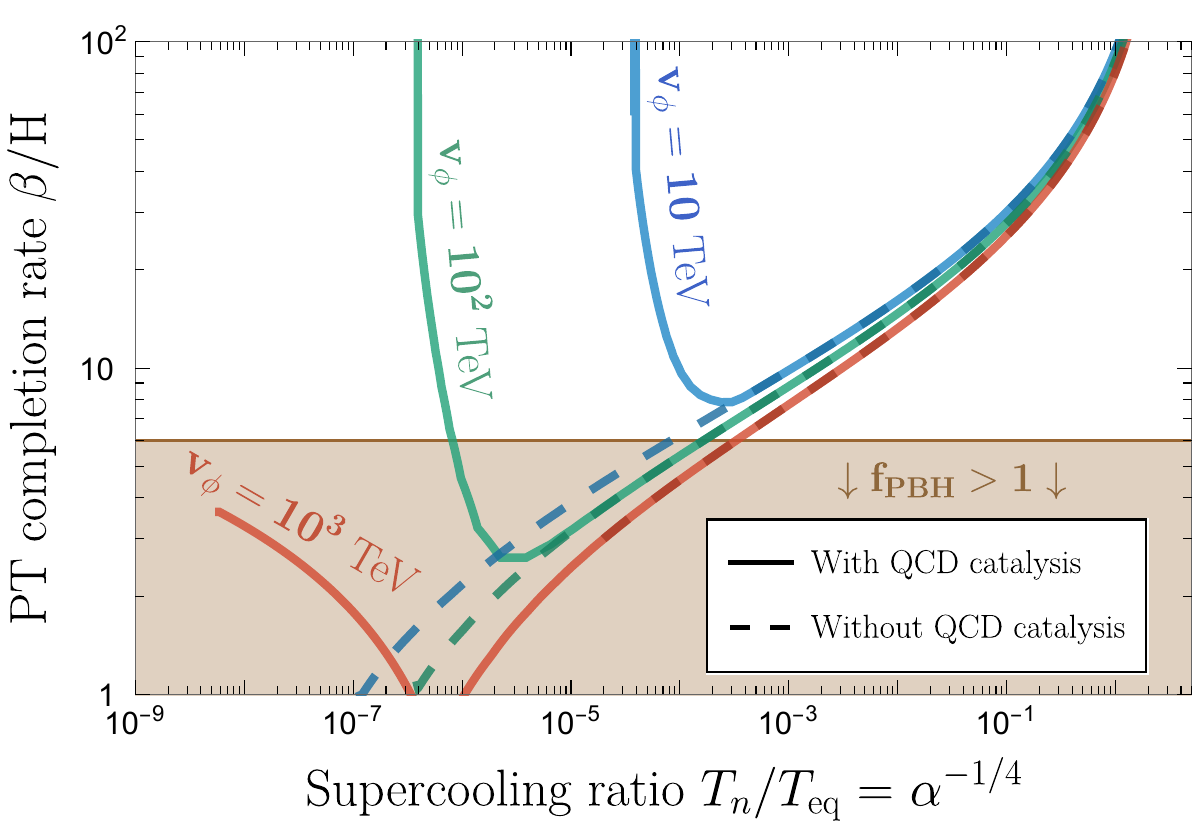}
\caption{ The PT completion rate $\beta/H$ becomes lower as the amount of supercooling increases, until QCD effects kick in around $T_n\simeq T_{\mathsmaller{\rm QCD}}$ in Eq.~\eqref{eq:S3overT_cw}. The lines are obtained from varying $g_{\rm D}$. The parameter $\alpha$ along the horizontal axis is the  latent heat fraction.}
\label{fig:beta_vs_Tn}
\end{figure}

We can now evaluate the rate of PT completion $\beta \equiv d\log{\Gamma_{\mathsmaller{\rm V}}}/dt$ at nucleation.
Above QCD confinement $T_n\gg T_{\mathsmaller{\rm QCD}}$, it decreases logarithmically with the temperature
\begin{equation}
\beta/ H_n \simeq -4 + \frac{127}{g_{\rm D}^3 \ln^2{\left( \frac{M}{T_n}\right)}} .
\end{equation}
Instead, around $T_n\sim T_{\mathsmaller{\rm QCD}}$ the PT completion rate diverges as $\beta/ H_n \propto 1/( T_{\mathsmaller{\rm QCD}}-T_n)$. We conclude that the PT rate $\beta/ H$ is minimized for coupling $g_{\rm D}$ associated with the longest supercooling  stage while not being catalyzed by QCD  $\Lambda_{\rm QCD} \lesssim T_n\ll T_{\rm eq}$. The nucleation temperature $T_n$ and the PT rate $\beta/H_n$ in scale-invariant SM+$U(1)_{\rm D}$ are shown in Fig.~\ref{fig:beta_vs_Tn}.

The latent heat $\Delta V \simeq \beta_\lambda \vphi^4/16$ dominates the energy density of the universe, generating an inflationary period, below the temperature $T_{\rm eq}$:
\begin{equation}
T_{\rm eq}  \simeq \frac{g_{\rm D}\vphi}{10.85} \left( \frac{100}{g_*}\right)^{1/4}.
\end{equation}

\underline{Reheating after matter domination} ---
Supercooled PTs proceed with ultra-relativistic bubble wall Lorentz factor $\gamma \gg 1$ \cite{Bodeker:2017cim,Azatov:2020ufh,Gouttenoire:2021kjv,Azatov:2023xem}.
Bubble growth convert the latent heat into scalar field gradient stored in the walls and plasma excitation resulting from the friction operating on the walls. The fraction stored in the former at percolation is given by:
\begin{equation}
\label{eq:kappa_coll}
\kappa_{\rm coll} = \frac{\gamma_{\rm coll}}{\gamma_{\rm max}},
\end{equation}
where $\gamma_{\rm coll}$ is the wall Lorentz factor at collision accounting for friction \cite{Gouttenoire:2021kjv}, while $\gamma_{\rm max}$ is the same quantity neglecting friction. We refer to App.~\ref{app:GW} for the details. The main dark Higgs decay channel is into two Higgs $\phi \to hh$. However the decay rate is suppressed by the small mixing $\lambda_{h\phi}$:
\begin{equation}
\label{eq:Gamma_phi_hh}
    \Gamma_{\phi\to hh} = \frac{\lambda_{h\phi}^2 v_\phi^2}{32\pi m_\phi} = 0.3H_n \left( \frac{0.5}{g_{\rm D}}\right)^6 \left( \frac{10^3~\rm TeV}{v_{\phi}}\right)^5,
\end{equation}
where $m_\phi =\beta_\lambda v_\phi$ is the scalar mass and $H_n^2=\Delta V/3M_{\rm pl}^2$ is the Hubble rate at percolation.
The universe after percolation contains a matter component occupying a fraction $\kappa_{\rm coll}$ in Eq.~\eqref{eq:kappa_coll} of the total energy density. According to Eq.~\eqref{eq:Gamma_phi_hh}, this matter component is long-lived for $v_\phi \gtrsim 10^3~\rm TeV$. It dominates the universe below the temperature $T_{\rm dom} = \kappa_{\rm coll}T_{\rm eq}$, hence generating an early matter-domination (MD) era, which ends with the decay of $\phi$ at the temperature $ T_{\rm dec} = 1.2\sqrt{M_{\rm pl} \Gamma_{\phi}}/{g_*^{1/4}}$.
The decay of $\phi$ reheates the SM and dilutes the abundance of any relics decoupled from the SM with a factor:
\begin{equation}
\label{eq:dil_fac_D}
    D= 1+\frac{T_{\rm dom}}{T_{\rm dec}}.
\end{equation}
As discussed in the next section, this will have a minor impact on the PBH detectability but a major impact on the GW one.

\begin{figure}[h!]
\centering
\includegraphics[width=240pt]{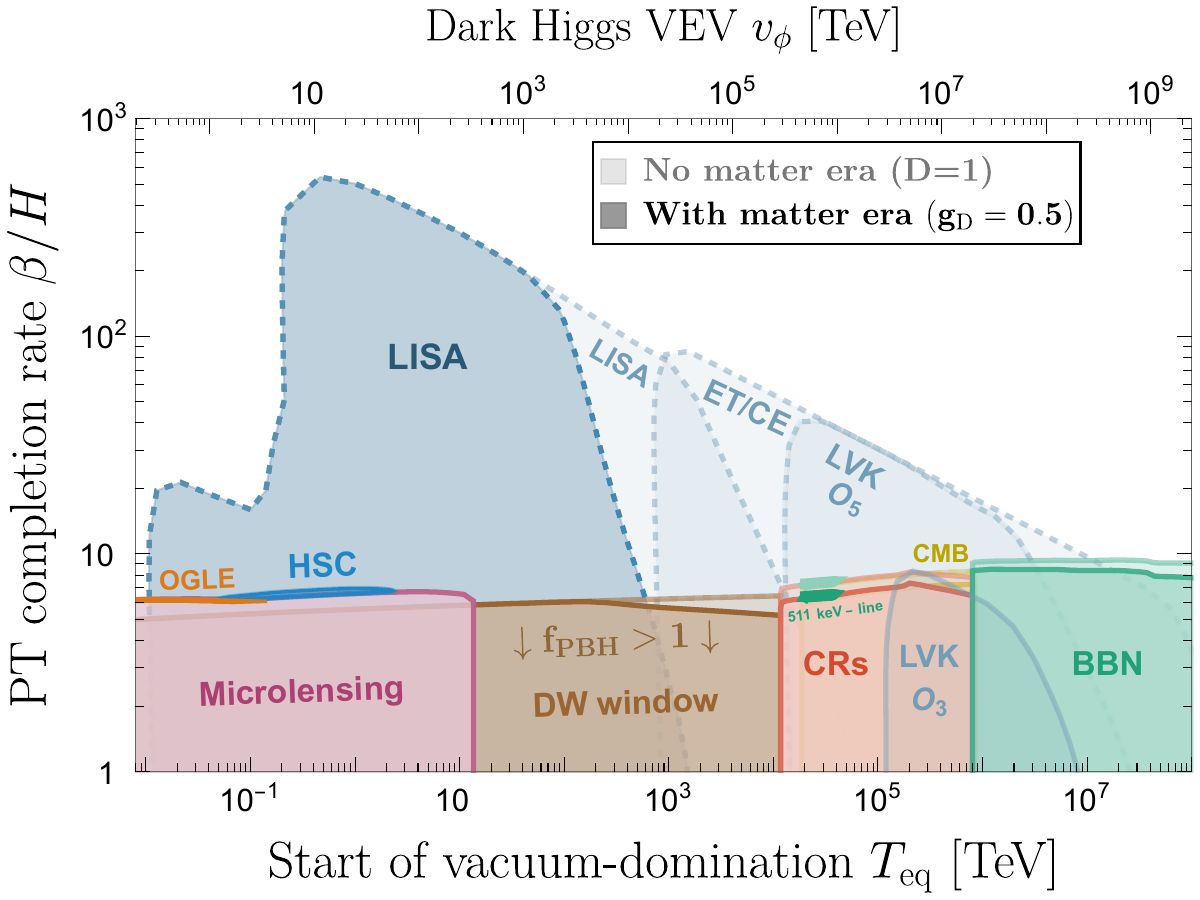}
\caption{ In an adiabatic universe, supercooled phase transition produce gravitational waves within the frequency window of future interferometers LVK run O3 \cite{KAGRA:2021kbb}, run $O_5$ \cite{LIGOScientific:2014pky}, ET/CE \cite{Punturo:2010zz,Maggiore:2019uih,Reitze:2019iox} and LISA \cite{Audley:2017drz,Robson:2018ifk,LISACosmologyWorkingGroup:2022jok,Smith:2019wny} with an amplitude larger than astrophysical foregrounds from white dwarfs binaries and compact binaries \cite{Lamberts:2019nyk,Boileau:2021gbr,Robson:2018ifk,Rosado:2011kv,KAGRA:2021kbb}. Instead, in the minimal scale-invariant extension of the SM, the long lifetime of the scalar field dilute all the GW signals at large dark Higgs VEV $v_{\phi}\gtrsim 10^4~\rm TeV$. Instead the PBH abundance is only slighly impacted by the matter era. See App.~\ref{app:GW} for more details on the experiments reach and expected astrophysical foregrounds.}
\label{fig:GW_spectra_sens_betaOH}
\end{figure}

\begin{figure*}[ht!]
\centering
\includegraphics[width=500pt]{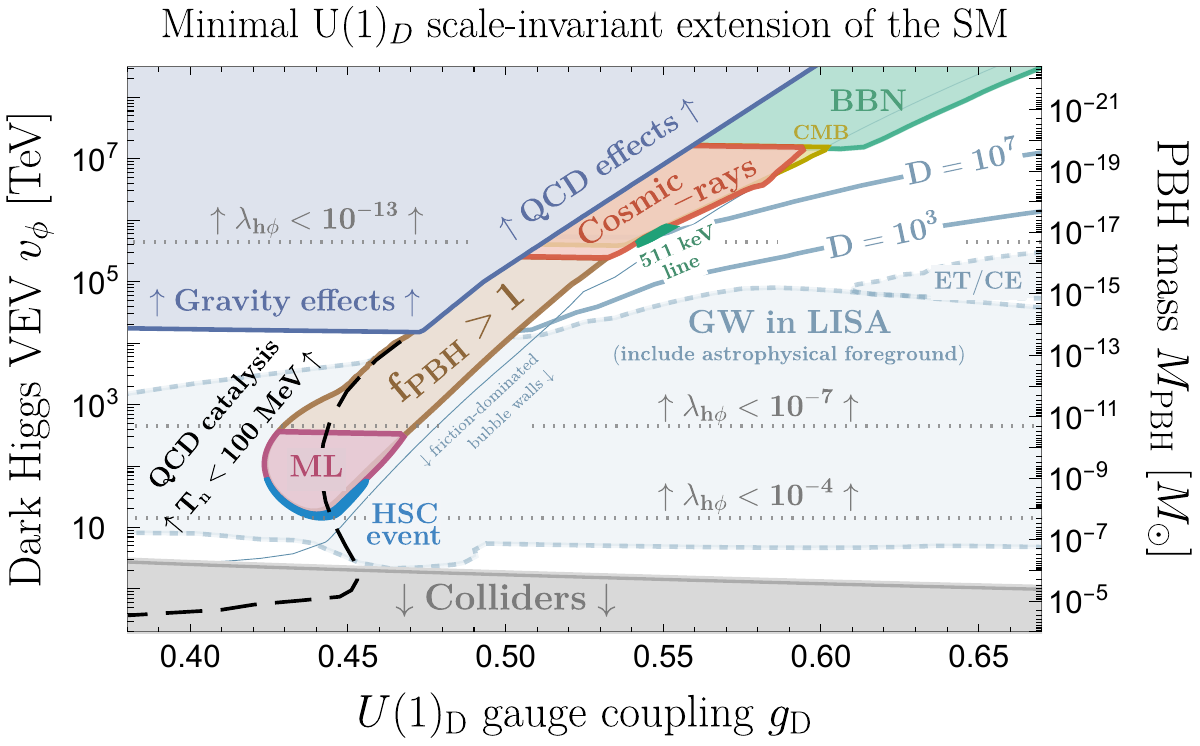}
\caption{ PBH and GW predictions in the minimal scale-invariant $U(1)_{\rm D}$ extension of the SM. The model has only two parameters: the hidden $U(1)_{\rm D}$ breaking scale $\vphi$ and its gauge coupling $g_{\rm D}$. 
The purple region shows the microlensing constraints (ML) from HSC/subaru telecope \cite{Smyth:2019whb,Sugiyama:2019dgt}. The blue U-shaped region can explain the HSC event \cite{Niikura:2019kqi,Sugiyama:2021xqg,Kusenko:2020pcg}. The boundary of the \textbf{brown} region can explain dark matter. Hawking evaporation overproduce cosmic rays 
 in the \textbf{red} region \cite{DeRocco:2019fjq,Laha:2019ssq,Keith:2021guq,Carr:2009jm,Boudaud:2018hqb,Laha:2020ivk,Dasgupta:2019cae,Coogan:2020tuf,Korwar:2023kpy}, but could contribute to the $511~\rm keV$ excess in the green spot \cite{DeRocco:2019fjq,Laha:2019ssq,Keith:2021guq}. Non-observation of energy injection from Hawking radiation in CMB \cite{Poulin:2016anj,Stocker:2018avm,Poulter:2019ooo} and BBN \cite{Carr:2009jm,Acharya:2020jbv,Keith:2020jww} exclude the \textbf{yellow} and \textbf{green} regions. See \cite{Carr:2020gox} for a review of PBH constraints.
 The GW constraints shown in \textbf{light blue} disappear at large $v_{\phi}\gtrsim 10^4~\rm TeV$ due to entropy injection following the decay of the long-lived dark Higgs. The associated dilution factor $D$ is  indicated with blue lines labelled ``D''. Part of the PBH DM region can not be probed with GWs.
 On the left of the \textbf{dashed black} line, the PT is catalysed by QCD confinement. In the \textbf{blue} region in the top-left corner, the universe can not tunnel without ``QCD effects'', due to $\Gamma_{\mathsmaller{\rm V}} = H^4$ having no solution $T_n$. However, the QCD catalysis temperature given in Eq.~\eqref{eq:S3overT_cw} is smaller than the Hubble scale $T_{\mathsmaller{\rm QCD}}\lesssim H_n$, implying that the tunneling dynamics receives large correction from ``gravity effects'' \cite{Coleman:1980aw,Hawking:1981fz}, which we leave for future research.  The \textbf{gray} region displays collider constraints on additional Higgs boson \cite{ALEPH:2006tnd,Ilnicka:2018def,Robens:2015gla,Robens:2019kga}, see Fig.~\ref{fig:collider_constraints} in App.~\ref{app:nearly_conf_sec} for more details. On the right of the very \textbf{thin blue} line, bubble wall friction convert most of the the latent heat into the plasma, which reduces the energy fraction stored in the scalar field and therefore reduces the dilution factor $D$.}
\label{fig:PBHDM_vphi_gD_withQCD}
\end{figure*}

\section{PHENOMENOLOGY}

\underline{Primordial black holes} ---
Due to the stochastic nature of thermal tunneling, distinct causal patches percolate at slightly different times. Patches which percolate later hold the latent heat longer in the form of vacuum energy while their neighbors already released it into radiation-like energy density. When the latest patches finally percolate, they are overdense which, beyond a given energy density threshold $\delta_c$, lead to their collapse into PBHs. 
We set the over-density threshold to $\delta_c = 0.45$, which assumes a Mexican hat density profile during radiation domination \cite{Jedamzik:1999am,Green:2004wb,Musco:2004ak,Harada:2015yda}. Different profile shapes would result in different values within the range $\delta_c \in [0.40,~0.67]$ \cite{Musco:2018rwt,Escriva:2019phb}. The determination of the density profile shape as well as the precise equation of state right after percolation are left for future studies. In particular, in the runaway bubble wall regime with $\kappa_{\rm coll}=1$, PBH formation could occur during a matter era with a significantly lower critical threshold \cite{Harada:2016mhb}, leading to an enhanced PBH abundance. 
The PBH mass is given by the mass within the sound horizon at the time of percolation:
\begin{align}
\label{eq:MPBHs}
M_{\rm \mathsmaller{PBH}} \simeq M_{\mathMoon}~\left( \frac{0.5}{g_{\rm D}}\right)^2\left(\frac{\lambda_{h\phi}}{1.2\times 10^{-4}} \right),
\end{align}
where $M_{\mathMoon} \simeq 3.7\times 10^{-8} M_{\odot}$ is the moon mass.
{The calculation of the PBH abundance has been completed in the companion paper~\cite{Gouttenoire:2023naa} and reported in App.~\ref{app:PBH_formation}.} Normalised to the dark matter relic density, it gives:
\begin{equation}
\label{eq:f_PBH_main}
f_{\rm \mathsmaller{PBH}} \simeq \frac{1}{D}\left(\frac{\mathcal{P}_{\rm coll}}{ 3.1 \times 10^{-12}}\right)\left( \frac{T_{\rm eq}}{1~\rm TeV} \right)\,.
\end{equation}
where $\mathcal{P}_{\rm coll}$ is the fraction of causal patches collapsing into PBHs, see App.~\ref{app:PBH_formation}. We added the dilution factor $D$ in Eq.~\eqref{eq:dil_fac_D} to account for the entropy injection resulting from the matter era after the PT. In Fig.~\ref{fig:GW_spectra_sens_betaOH}, we show that supercooled PTs with $\beta/H\lesssim 7$ would produce PBHs detectable by on-going experiments, and that the effect of the dilution factor $D$ -- opaque vs transparent -- is very limited. PBHs formation in the scale-invariant $U(1)_{\rm D}$ extension of the SM is shown in Fig.~\ref{fig:PBHDM_vphi_gD_withQCD} together with astrophysical and cosmological constraints \cite{Carr:2020gox}.  For comparison, in Fig.~\ref{fig:PBHDM_vphi_gD_noQCD} we present the same figure with QCD catalysis neglected. This is equivalent to setting $\left<h\right>_{\mathsmaller{\rm QCD}}$ in Eq.~\eqref{eq:h_QCD_formula} to $0$ instead of $100~\rm MeV$. 

\underline{Gravitational waves} ---
 As the supercooled PT completes, the large latent heat $\Delta V$ is converted into scalar field gradients with an energy fraction $\kappa_{\rm coll}$. This drives bubble walls to ultra-relativistic Lorentz factors \cite{Bodeker:2017cim, Azatov:2020ufh, Gouttenoire:2021kjv, Azatov:2023xem}. A fraction $1-\kappa_{\rm coll}$ is converted into extremely thin shells of ultra-relativistic particles \cite{Baldes:2020kam, Azatov:2020ufh, Gouttenoire:2021kjv, Jinno:2022fom, Baldes:2023fsp, Azatov:2023xem} or ultra-relativistic shock-waves \cite{Espinosa:2010hh, Jinno:2019jhi, Cutting:2019zws, Lewicki:2022pdb}, which trail bubble walls \cite{Baldes:2020kam, Azatov:2020ufh, Gouttenoire:2021kjv, Jinno:2022fom, Baldes:2023fsp, Azatov:2023xem}. The GW spectrum from thin shells of stress-energy momentum tensor can be described by the ``bulk flow'' model \cite{Jinno:2017fby,Konstandin:2017sat, Lewicki:2020jiv, Lewicki:2020azd, Cutting:2020nla}. We refer the reader to App.~\ref{app:GW} for the details.
For large VEV $v_\phi \gtrsim 10^3~
\rm TeV$, corresponding to small Higgs mixing $\lambda_{h\phi}\lesssim 10^{-5}$, the PT is followed by a matter domination era, further redshifting the GW signal according to:
\begin{equation}
\Omega_{\rm GW}(f)\to \frac{\Omega_{\rm GW}(D^{1/3}f)}{D^{4/3}},
\end{equation}
where $D$ is given by Eq.~\eqref{eq:dil_fac_D} and with additional spectral distortion in the causality tail \cite{Barenboim:2016mjm,Domenech:2020kqm,Ellis:2020nnr,Hook:2020phx,Ertas:2021xeh}. We refer to Fig.~\ref{fig:GW_spectra_sens} in App.~\ref{app:GW} for more details and for a visualisation of the GW spectra. The impact of the dilution factor on the GW detectability is visible in Figs.~\ref{fig:GW_spectra_sens_betaOH} and \ref{fig:PBHDM_vphi_gD_withQCD}. We can see that the existence of the matter era -- which is a consequence of the smallness of the Higgs mixing $\lambda_{h\phi}\ll 1$ in Eq.~\eqref{eq:Higgs_mixing} -- limits the GW detectability to $v_{\rm \phi}\lesssim 10^4~\rm TeV$, corresponding to PBH masses $M_{\rm PBH}\gtrsim 10^{-14}~M_{\odot}$. Part of the region explaining DM with PBHs might never be probed with GWs, in contrast to the naive expectation \cite{Bartolo:2018evs}.x As a comparison, the same figure with matter era effects neglected is provided in Fig.~\ref{fig:CL_Tn_beta} in App.~\ref{app:nearly_conf_sec}.

\section{CONCLUSION}

{In this \textit{letter}, we have demonstrated that particle physics models featuring scalar potentials dominated by logarithmic terms, similar to the Coleman-Weinberg potential in Eq.~\eqref{eq:CW_potential_intro}, are promising theoretical frameworks for producing PBHs from 1stOPTs. By focusing on the $U(1)_{\rm D}$ scale-invariant extension of electroweak theory, we have found that PBHs with lunar masses can be produced for dark Higgs vev $v_\phi \sim 20~\rm TeV$. This provides a plausible interpretation for the candidate event from the HSC microlensing search \cite{Niikura:2017zjd,Sugiyama:2021xqg,Kusenko:2020pcg}.} At larger vev $v_\phi \in 4\times [10^2,~10^5]~\rm TeV$, PBHs can explain $100\%$ of the observed Dark Matter (DM) relic density \cite{Carr:2016drx,Carr:2020xqk,Carr:2020gox,Green:2020jor}. 
Another consequence of the larger hierarchy of scales is the suppression of the dark Higgs decay width as $\Gamma_{\rm \phi}\propto (v_{\rm EW}/v_\phi)^4$. The decay after a matter era can generate a tremendous amount of entropy dump which dilutes the gravitational wave (GW) signal. GW constraints disappear for $v_{\phi}\gtrsim 10^{4}~\rm TeV$, leaving PBHs as the sole detectable evidence for a scale-invariant extension of the standard model across a wide range of energy scales. This includes areas that could account for 100$\%$ of DM in terms of PBHs. As pointed in \cite{Ray:2021mxu,Ghosh:2021gfa,Keith:2022sow,Malyshev:2022uju,Malyshev:2023oox}, part of the PBH DM region will be explored with future telescopes \cite{e-ASTROGAM:2017pxr,AMEGO:2019gny,Labanti:2021gji}. The detection of PBH dark matter without GW counterpart could be a smoking-gun for a conformal Higgs scenario. However, it would be degenerate with standard PBH formation scenarios if they are followed by entropy injection. This degeneracy could be resolved by measuring the mass distribution of PBHs \cite{Baldes:2023rqv,Lewicki:2024ghw} or by using cosmic string archaeology to decipher the equation of state, duration and epoch of the entropy injection phase \cite{Cui:2017ufi,Cui:2018rwi,Gouttenoire:2019rtn,Gouttenoire:2019kij,Gouttenoire:2021wzu,Gouttenoire:2021jhk,Ghoshal:2023sfa}.
{Two remarks are important to highlight before closing. Firstly, in contrast to logarithmic potentials, traditional polynomial potentials typically do not satisfy the two necessary conditions for PBH formation which are latent heat domination ($T_n \ll T_c$) and slow rate of completion ($\beta/H \lesssim 7$). This is because the bounce action decreases too rapidly with temperature, following a power-law \cite{Dine:1992wr} instead of the logarithmic decrease in Eq.~\eqref{eq:S3overT_cw}.
Secondly, it is important to emphasize that logarithmic potentials are not exclusively predicted by weakly-coupled Coleman-Weinberg models. They also arise in nearly-conformal strongly-coupled models featuring a light dilaton~\cite{Arunasalam:2017ajm,Bruggisser:2018mus, Bruggisser:2018mrt, Baratella:2018pxi, Agashe:2019lhy, DelleRose:2019pgi, VonHarling:2019rgb,Bloch:2019bvc,Baldes:2020kam,Baldes:2021aph} and their weakly-coupled 5D holographic dual~\cite{Randall:1999ee,Randall:1999vf,Goldberger:1999uk,Creminelli:2001th,Randall:2006py,Nardini:2007me,Hassanain:2007js, Konstandin:2010cd, Konstandin:2011dr, vonHarling:2017yew,Fujikura:2019oyi,Bunk:2017fic, Dillon:2017ctw, Megias:2018sxv,Megias:2020vek, Agashe:2020lfz}. This {\textit{letter}} opens new avenues for investigating the existence of PBHs in our universe using nearly scale-invariant particles physics models.}

{\bf Acknowledgements.}---%
The author thanks Iason Baldes, Brando Bellazzini, Raffaele Tito D'Agnolo, Ryusuke Jinno, Alexander Kusenko, Filippo Sala, Geraldine Servant,  Sunao Sugiyama, Bogumila Swiezewska, Carlo Tasillo, and Tomer Volansky for useful discussions. The author thanks Sunao Sugiyama and Tomer Volansky for feedback on a preliminary version of the manuscript. The author thanks Sunao Sugiyama for generously sharing the posterior distribution of the PBH interpretation of HSC and OGLE events.
Finally, the author is grateful to the Azrieli Foundation for the award of an Azrieli Fellowship.

\clearpage
\appendix
\onecolumngrid

\fontsize{11}{13}\selectfont






\makeatletter
\renewcommand*{\fnum@figure}{{\normalfont \normalsize \figurename~\thefigure}}
\renewcommand*{\@caption@fignum@sep}{ : }
\makeatother

\renewcommand{\tocname}{\Large  Table of contents
\vspace{1 cm}}%

\titleformat{\section}
{\normalfont\fontsize{12}{14}\bfseries  \centering }{\thesection.}{1em}{}
\titleformat{\subsection}
{\normalfont\fontsize{12}{14}\bfseries \centering}{\thesubsection.}{1em}{}
\titleformat{\subsubsection}
{\normalfont\fontsize{12}{14}\bfseries \centering}{\thesubsubsection)}{1em}{}

\titleformat{\paragraph}
{\normalfont\fontsize{12}{14}\bfseries  }{\thesection:}{1em}{}

 {
 \hypersetup{linkcolor=black}
 \tableofcontents
 }
\newpage
\section{Late-blooming mechanism}
\label{app:PBH_formation}

We report here the derivation of the PBH abundance during supercooled PTs which has been proposed {in the companion paper} \cite{Gouttenoire:2023naa}. It accounts for fluctuation in the nucleation time of the first bubble in the full past light-cone of the collapsing patch and consider the collapse to occur after percolation, in a radiation-dominated universe. In contrast, other approaches proposed in Refs.~\cite{Kodama:1982sf,Lewicki:2023ioy,Kawana:2022olo,Flores:2024lng} are restricting the collapsing patch to remain $100\%$ vacuum dominated until collapse, leading to a lower PBH abundance, while in Ref.~\cite{Liu:2021svg} nucleation is not accounted in the entire past light-cone of the collapsing patch, leading to a larger PBH abundance.
The derivation in the companion paper \cite{Gouttenoire:2023naa} has been reproduced in details in Ref.~\cite{Baldes:2023rqv}.
While in the companion paper \cite{Gouttenoire:2023naa} a monochromatic mass distribution is considered, Ref.~\cite{Baldes:2023rqv} extends the analysis by calculating the PBH mass distribution. Ref.~\cite{Lewicki:2024ghw}, which appeared after the first version of this work, improves the treatment of \cite{Gouttenoire:2023naa,Baldes:2023rqv} by accounting for fluctuations in the nucleation time of the first $j_c$ bubbles in a given causal patch, with $j_c$ being arbitrary. Refs.~\cite{Gouttenoire:2023naa,Baldes:2023rqv} can be viewed as the case $j_c=1$. We leave the quantitative analysis of the importance of those corrections to the PBH abundance for future works.

During a supercooled PT, the latent heat $\Delta V \simeq \beta_\lambda \vphi^4/16$ can dominate the energy density of the universe, leading to an inflationary stage below the temperature $T_{\rm eq}$:
\begin{equation}
\label{eq:Teq_def_2}
 \frac{\pi^2}{30}g_* T_{\rm eq}^4 \equiv \Delta V, 
\end{equation}
which also corresponds to the maximal reheating temperature after the PT, up to ratio of $g_*$.
Inflation ends when bubbles percolate around a temperature $T_n$ given by:
 \begin{equation}
\label{eq:inst_approx}
    \Gamma_{\mathsmaller{\rm V}} (T_{n}) = H^4(T_{n})\,.
\end{equation}
\begin{figure}[ht!]
\centering
\includegraphics[width=0.7\linewidth]{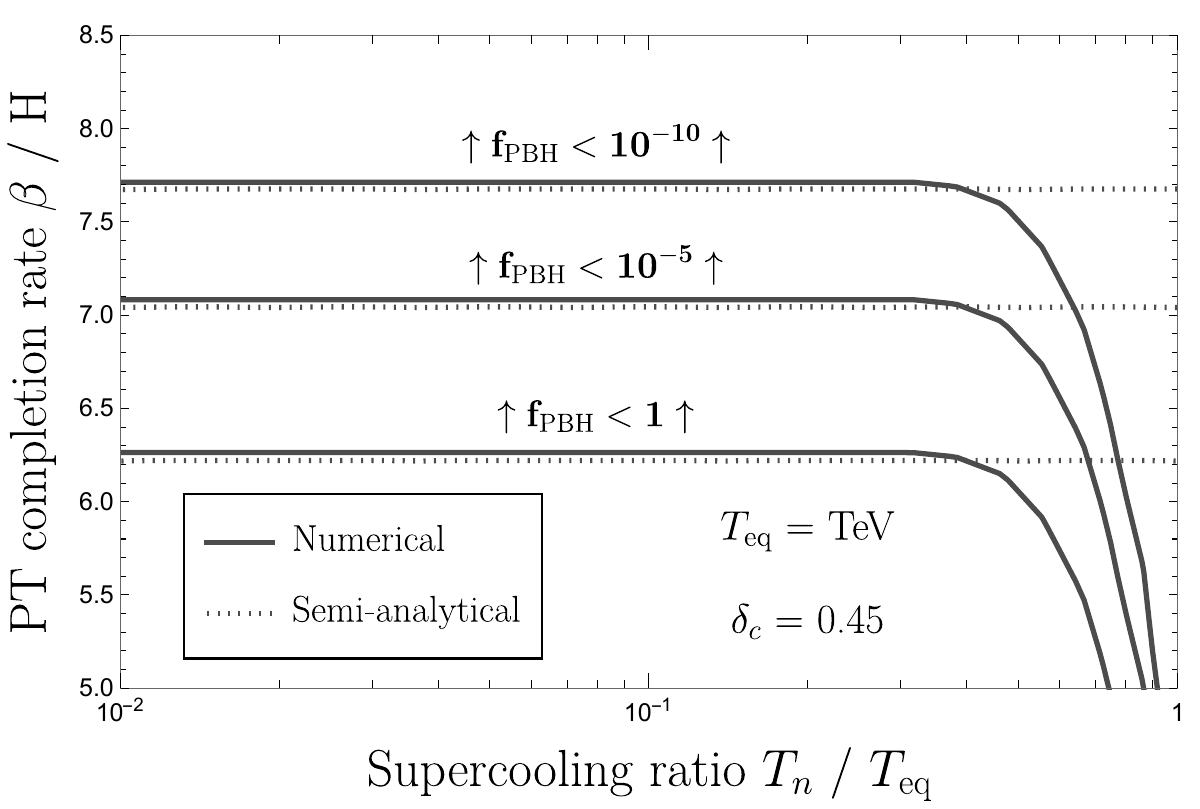}
\caption{\normalsize  PBH abundance as a function of the amount of supercooling $T_n/T_{\rm eq}$ and rate of completion $\beta/H$.}
\label{fig:f_PBH}
\end{figure}
It is convenient to approximate the tunneling rate with the bounce action Taylor-expanded around the nucleation time $t_n$ at first order:
\begin{equation}
\label{eq:tunneling_rate_def_0}
\Gamma_{\mathsmaller{\rm V}}(t) = \Gamma_0 e^{\beta t}\,\qquad \text{with} \quad \Gamma_0 = H^4(T_{n}) e^{-\beta t_n}\,,
\end{equation}
where the expression for $\Gamma_0$ follows from Eq.~\eqref{eq:inst_approx}.  The validity of the $\beta$-parametrization in Eq.~\eqref{eq:tunneling_rate_def_0} is discussed below Eq.~\eqref{eq:2ndorder_betaH}.

During the nucleation phase, the vacuum energy $\Delta V$ is converted into kinetic energy of bubble walls which redshifts akin radiation.
The time and position at which a bubble nucleates are stochastic variables. Hence the percolation time of each causal patch fluctuates with respect to the background. 
While the average universe has already percolated and is cooling like a radiation-dominated universe, late-bloomers -- patches which start nucleating later -- maintain an almost constant energy density until they finally percolate and convert their vacuum energy into radiation.
Hence, causal patches which nucleate the latest inflate the longest and become the densest.
 Late-blooming region (``late'') collapse into PBHs if the contrast radiation density with respect to the background (``bkg'') is larger than:
\begin{equation}
 \label{eq:PBH_threshold_0}
\delta(t;t_{n_i}) = \frac{\rho_{\rm R}^{\rm late}(t;t_{n_i})-\rho_{\rm R}^{\rm bkg}(t)}{\rho_{\rm R}^{\rm bkg}(t)}~ > ~\delta_{\rm c},
\end{equation}
where $t_{n_i}$ is the time at which a late-bloomer nucleates its first bubble. It is also the time until which it remains $100\%$ vacuum-dominated.
The overdensity generated by a supercooled 1stOPT is expected to source an adiabatic curvature perturbation $\zeta$ to the FLRW universe:
\begin{equation}
    ds^2 = -a^2dt^2 + a^2e^{2\zeta(r)}\left[dr^2 + r^2 d\Omega^2\right],
\end{equation}
where we have assumed a spherically-symmetry perturbation. Indeed, a connection between energy density contrast and curvature perturbation arises from the linearized Einstein equations \cite{Harada:2015yda,Young:2019yug,Franciolini:2022tfm}:
\begin{equation}
\frac{\rho_R^{\rm late}(r,t;t_{n_i}) - \rho_R^{\rm bkg}(r,t)}{\rho_R^{\rm bkg}(r,t)} \simeq - \frac{8}{9}\left(\frac{1}{aH}\right)^2 e^{-5\zeta(r)/2} \Delta \left(e^{\zeta(r)/2}\right).
\label{eq:eqn:non-linear}
\end{equation}
The density contrast in Eq.~\eqref{eq:PBH_threshold_0} is the density contrast in Eq.~\eqref{eq:eqn:non-linear} averaged over a Hubble sphere. This motivates to apply the results on PBH formation from curvature perturbations generated by inflation to the curvature generated during a 1stOPT.\footnote{This approach conflicts with the claims made by Ref.~\cite{Flores:2024lng}, which appeared after the first version of this work, and whose resolution is left for future works.}
Large curvature peaks collapse into PBHs if the averaged associated density contrast is larger than a given threshold value $\delta_c$ \cite{Shibata:1999zs}. For collapse occurring during radiation domination era, the threshold has been found to take values within the range $\delta_c \in [0.40,~0.67]$ depending on the profile shape \cite{Musco:2018rwt,Escriva:2019phb}. Specifically, $\delta_c = 0.40$ corresponds to the broadest profiles, while $\delta_c = 0.67$ corresponds to the sharpest profiles. Awaiting future studies to determine the shape of the density profile generated by 1stOPT, we assume $\delta_c = 0.45$, corresponding to a Mexican hat profile shape in radiation-domination. Such choice has been widely used in the literature \cite{Jedamzik:1999am,Green:2004wb,Musco:2004ak,Harada:2015yda}.

In presence of bubble growth, the vacuum energy density decreases as:
\begin{equation}
\label{eq:rho_V_text}
\rho_{V}(t;t_{n_i}) =  F(t;\, t_{n_i})\Delta V\,,
\end{equation}
where $F(t;\,t_{n_i})$ is the volume fraction of remaining false vacuum at time $t$ \cite{Guth:1979bh}:
\begin{equation}
\label{eq:F_vacuum_fraction_text}
F(t;\, t_{n_i}) = \textrm{exp}\hspace{-0.05cm}\left[-\hspace{-0.1cm}\int_{t_{n_i}}^t \hspace{-0.2cm}dt{'}\, \Gamma_{\mathsmaller{\rm V}}(t^{'}) a(t{'})^3 \frac{4\pi}{3} \left(\int_{t^{'}}^t\frac{d\tilde{t}}{a(\tilde{t})}\right)^{\!\!3}\right],
\end{equation}
where $a(t)$ is the scale factor of the universe evolving with energy density $\rho_{\rm tot}=\rho_{\rm V}+\rho_{\rm R}$.
Energy-conservation in an expanding universe implies:
\begin{equation}
\label{eq:rho_R_text}
\dot{\rho}_{\rm R}(t;t_{n_i})  + 4H\rho_{\rm R}(t;t_{n_i}) = - \dot{\rho}_{V}(t;t_{n_i}) .
\end{equation}
The critical time of nucleation postponing $t_{n_i}^{\rm PBH}$ beyond which a causal patch reaches the threshold in Eq.~\eqref{eq:PBH_threshold_0} can be found from numerically solving for $\rho_{\rm R}(t;t_{n_i})$ in Eq.~\eqref{eq:rho_R_text} as a function of $t_{n_i}$ for late-bloomers and at $t_{n_i} \to 0$ for the background.
The fraction $\mathcal{P}_{\rm coll}$ of causal patches which collapse into PBHs is given by the probability that causal patches remain vacuum-dominated until $t_{n_i}^{\rm PBH}$.  Denoting by $t_{\rm max}$ the time when the density contrast reaches its maximal value, set equal to the threshold in Eq.~\eqref{eq:PBH_threshold_0}, we have: 
\begin{equation}
\label{eq:proba_tni_PBHs_0}
\mathcal{P}_{\rm coll}  = \exp\hspace{-0.1cm}\left[ -\hspace{-0.1cm}\int_{0}^{t_{n_i}^{\rm PBH}} \hspace{-0.2cm}dt{'}\, \Gamma_{\mathsmaller{\rm V}}(t{'}) a(t{'})^3V(t';t_{\rm max}) \right]\,,
\end{equation}
where $V(t'; t_{\rm max})$ is the comoving volume at time $t'$ which is in causal contact with the collapsing patch at $t_{\rm max}$:
\begin{equation}
\label{eq:causal_volume_text}
    V(t';t_{\rm max})  = \frac{4\pi}{3}\left(r_{\rm H}(t_{\rm max}) + r(t_{\rm max};t{'}) \right)^3\,,
\end{equation}
with $r_{\rm H}\equiv (aH)^{-1}$ the comoving horizon and $r(t;t')= \int_{t^{'}}^td\tilde{t}/a(\tilde{t})$ the light travel distance between $t'$ and $t$. 

In Ref.~\cite{Gouttenoire:2023naa}, a ready-to-use formula fitted on the numerical calculation is proposed, valid in the supercooled limit $T_n\ll T_{\rm eq}$:
\begin{equation}
\label{eq:proba_coll_ana}
\mathcal{P}_{\rm coll}\simeq \exp\left[-a\left( \frac{\beta}{H_n} \right)^{\!b}\left(1+\delta_{\rm c}\right)^{c\frac{\beta}{H_n}} \right]\,,
\end{equation}
with $a\simeq 0.5646$, $b\simeq 1.266$, $c\simeq 0.6639$ and $\delta_c\simeq 0.45$. 
Since the collapse occurs during radiation, the PBH mass is given by the mass inside the sound horizon $c_s H^{-1}$ at the time of the collapse, e.g. \cite{Escriva:2021pmf}:
\begin{align}
\label{eq:MPBHs}
M_{\rm \mathsmaller{PBH}} \simeq M_{\mathMoon}~\left(\frac{106.75}{g_*(T_{\rm eq})} \right)^{1/2}\left(\frac{500~\rm GeV}{T_{\rm eq}} \right)^2,
\end{align}
where $c_s =1/\sqrt{3}$ was used and where $M_{\mathMoon} \simeq 3.7\times 10^{-8} M_{\odot}$ is the moon mass. We refer to \cite{Lewicki:2024ghw}.
Counting the number of collapsing patches in our past light-cone and weighting them by $M_{\rm \mathsmaller{PBH}}$, we obtain the PBH abundance in unit of the DM relic density:
\begin{equation}
\label{eq:f_PBH_main}
f_{\rm \mathsmaller{PBH}} \simeq \left(\frac{\mathcal{P}_{\rm coll}}{ 3.1 \times 10^{-12}}\right)\left( \frac{T_{\rm eq}}{1~\rm TeV} \right)\,.
\end{equation}
As we show in Fig.~\ref{fig:f_PBH}, values of $\beta/H\lesssim 7$ are needed to produce PBHs in observable abundance $f_{\rm \mathsmaller{PBH}}\gtrsim 10^{-5}$.

\section{Scale-invariant SM extension }
\label{app:nearly_conf_sec}
\subsection{SM only}
In the SM, the electroweak scale can not be explained by radiative correction alone and a mass scale must be added at tree level. This is because the one-loop contribution to the Higgs potential reads \cite{Coleman:1973jx}:
\begin{equation}
V(h,T=0) = \frac{\beta_{\lambda_h}}{4} h^4 \left(\ln{\left( \frac{h}{v_{\rm EW}}\right)} - \frac{1}{4}\right),\qquad (4\pi)^2\beta_{\lambda_h}=\frac{6m_W^4+3m_Z^4-12m_t^4}{v_{\rm EW}^4} = -0.035.
\end{equation}
Due to the large top mass, the constant  $\beta_{\lambda_h} <0$ has the wrong sign and $h=0$ is the only minima of the potential. 
\subsection{Effective potential}
To generate the weak scale with a scale-invariant set-up, we complete the SM with a complex scalar $\Phi=\phi e^{\varphi/\sqrt{2}}$ charged under $U(1)_{\rm D}$. The tree level potential in the unitarity gauge reads:
\begin{equation}
V_{\rm tree}(h,\phi) = \frac{\lambda_h}{4}h^4 + \frac{\lambda_\phi}{4}\phi^4 - \frac{\lambda_{h\phi}}{4}h^2 \phi^2.
\end{equation} 
The Renormalized Group Equations (RGE) of the theory are given by, e.g. \cite{Khoze:2014xha}:
\begin{align}
&(4\pi)^2\frac{d g_{\rm D}}{d\log{\phi}} = g_{\rm D}^3/3 , \\
\label{eq:RGE_lambda_h_phi_app}
&(4\pi)^2\frac{d \lambda_{\rm h\phi}}{d\log{\phi}} = \lambda_{h\phi}\left(6y_t^2+12\lambda_h +8\lambda_\phi - 4\lambda_{\rm h\phi} - 6g_{\rm D}^2 \right) , \\
\label{eq:RGE_lambda_phi_app}
&(4\pi)^2\frac{d\lambda_\phi}{d\log{\phi}} =  6g_{\rm D}^4 + 20\lambda_{\phi}^2 + 2\lambda_{hs}^2  - 12 \lambda_{\phi} g_{\rm D}^2 . 
\end{align}
Integrating the first equation, we obtain:
\begin{equation}
\label{eq:gD_phi}
g_{\rm D}^2(\phi) = \frac{g_{\rm D}^2(\vphi) }{1-\frac{g_{\rm D}^2(\vphi)}{24\pi^2} t}, \qquad t \equiv \log\left(\frac{\phi}{\vphi}\right).
\end{equation}
The field excursion $\phi_*$ during tunneling is of the order of the nucleation temperature $\phi_* \sim T_n$, see Eq.~\eqref{eq:exit_point} below. The interval $1\lesssim T_n/\vphi \lesssim 10^{-10}$ characteristic of supercooled phase transition implies $-23\lesssim t \lesssim 0$. Hence, we can safely neglect the running of $g_{\rm D}$:
\begin{equation}
t \ll \frac{24\pi^2}{g_{\rm D}^2} \simeq 950\left( \frac{0.5}{g_{\rm D}}\right)^2 \qquad \implies \qquad  g_{\rm D}(\phi) \simeq g_{\rm D}(\vphi) .
\end{equation}
Due to collider bounds, the Higgs-mixing is small $\lambda_{h\phi}\lesssim 10^{-2}$, so we can safely neglect its RGE running in Eq.~\eqref{eq:RGE_lambda_h_phi_app}. The integration of Eq.~\eqref{eq:RGE_lambda_phi_app} gives:
\begin{equation}
\label{eq:lambda_phi}
\lambda_{\phi}(\phi) = \frac{3g_{\rm D}^2}{10} +\frac{\sqrt{21g_{\rm D}^4+10\lambda_{h\phi}^2}}{10}\tan{\left\{ \frac{\sqrt{21g_{\rm D}^4+10\lambda_{h\phi}^2}}{8\pi^2} t - \arctan{\left[\frac{3g_{\rm D}^2-10 \lambda_\phi^0}{\sqrt{21g_{\rm D}^4+10\lambda_{h\phi}^2}}\right]} \right\}},
\end{equation}
where $\lambda_\phi^0 \equiv \lambda_\phi(\vphi)$ is a boundary value. 
The potential at 1-loop with counterterm $\delta \lambda \phi^4/4$ is:
\begin{equation}
\label{eq:V_1-loop_T_zero}
V_{\rm 1-loop}^{T=0}(\phi,h) = \frac{\lambda_h}{4} h^4 - \frac{\lambda_{h\phi}}{4} h^2\phi^2 + \delta\lambda\, \phi^4 + \frac{\lambda_{\phi}(\phi)}{4}\phi^4.
\end{equation}
We impose the renormalization condition that $\phi = \vphi$ is the minimum of the potential:
\begin{equation}
\frac{\partial}{\partial \phi}\left[\delta \lambda\, \phi^4 + \frac{\lambda_\phi(\phi)}{4}\phi^4\right]_{\phi=\vphi} = 0,
\end{equation}
which leads to:
\begin{equation}
\label{eq:delta_lambda}
\delta \lambda = \frac{-3g_{\rm D}^4 -\lambda_{h\phi}^2+ 6g_{\rm D}^2 \lambda_\phi^0 -32 \pi^2 \lambda^0_\phi - 10 (\lambda_\phi^0)^2}{128\pi^2}.
\end{equation}
Plugging Eqs.~\eqref{eq:lambda_phi} and \eqref{eq:delta_lambda} into Eq.~\eqref{eq:V_1-loop_T_zero} and Taylor-expanding as a function of the logarithm $t$, we obtain:
\begin{equation}
V_{\rm 1-loop}^{T=0}(\phi,h) = \frac{\lambda_h}{4} h^4 - \frac{\lambda_{h\phi}}{2} h^2\phi^2 + \frac{\beta_\lambda}{4}\phi^4\left[\ln{\left(\frac{\phi}{\vphi} \right)}- \frac{1}{4} \right]+\frac{\kappa_\lambda}{4}\phi^4\ln^2{\left(\frac{\phi}{\vphi} \right)}+\cdots,
\end{equation}
with:
\begin{align}
&\beta_{\lambda} = \frac{1}{(4\pi)^2}\left[ 6g_{\rm D}^4 + 20(\lambda_{\phi}^0)^2 + 2\lambda_{h\phi}^2  - 12 \lambda_{\phi}^0 g_{\rm D}^2  \right] \simeq 6\alpha_{\rm D}^2, \\
&\kappa_{\lambda} = \frac{1}{(4\pi)^4}\left[ -36g_{\rm D}^6+192g_{\rm D}^4\lambda_\phi^0-360 g_{\rm D}^2 (\lambda_\phi^0)^2 + 400 (\lambda_{\phi}^0)^3 -12 g_{\rm D}^2 \lambda_{h\phi}^2 + 40 \lambda_\phi^0\lambda_{h\phi}^2\right] \simeq -\frac{9\alpha_{\rm D}^3}{\pi},
\end{align}
 where the two right-hand sides assume $g_{\rm D}^2\gtrsim \lambda_\phi^0,\lambda_{h\phi}$. We conclude that we can safely truncate the quantum loop correction to first-order in $t$ (leading-log approximation) as long as:
 \begin{equation}
     t~\ll~ \frac{8\pi^2}{3g_{\rm D}^2}  \simeq 105\left( \frac{0.5}{g_{\rm D}}\right)^2 \quad \implies \quad \kappa_\lambda \to 0.
 \end{equation}
 In this study, to minimize the number of parameters, we assume $\lambda_{\phi}^0 = \lambda_{\phi}(\vphi)\ll\sqrt{g_{\rm D}}$, which in practice leads us to set $\lambda_{\phi}^0=0$. Instead, despite its small value given in the main text, we keep $\lambda_{h\phi}$ finite in our equations.

Computing the second derivative of the potential around its minimum, we get the scalar mass which is loop-suppressed with respect to the vector boson mass:
\begin{equation}
m_\phi^2 = \beta_\lambda \vphi^2, \qquad \textrm{and} \qquad m_V^2 = g_{\rm D}^2 \vphi^2.
\end{equation}
The total potential is given by:
\begin{equation}
\label{eq:app_potential_full}
V(\phi) =  \frac{\lambda_{\phi}(\phi)}{4}\phi^4 + V_{\rm T}(\phi) + V_{\rm QCD}(\phi).
\end{equation}
The second piece contains the finite temperature correction at one loop and the Daisy contributions which accounts for the resummation of the Matsubara zero modes \cite{Dolan:1973qd}:
\begin{equation}
\label{eq:ThermalPotential_app}
 V_{\rm T}(\phi) =  \frac{ T^{4}}{2\pi^2} J_{B}\left(\frac{m_{V}^2(\phi)}{T^2}\right)+\frac{T}{12\pi} \left[ m_{V}^3 - \left( m_{V}^{2} + \Pi_{V}\right)^{3/2} \right] ,
\end{equation}
with vacuum and thermal masses $m_{V}^2(\phi)=g_{\rm D}^2 \phi^2$ and $\Pi_{V} =  g_{\rm D}^2 T^{2}/6$ respectively \cite{Carrington:1991hz}.
The third piece contains the Higgs contribution below the temperature of QCD confinement, see main text
\begin{align}
V_{\rm QCD}(\phi) = -\frac{1}{2}m_{\mathsmaller{\rm QCD}}^2\phi^2, \qquad \textrm{with} \quad m_{\mathsmaller{\rm QCD}}^2 = \frac{1}{2}\lambda_{h\phi}\left<h\right>_{\mathsmaller{\rm QCD}}^2\Theta(T-\Lambda_{\rm QCD}).
\end{align}
\subsection{Bounce action}
\label{app:bounce_action}
The cost for nucleating a bubble through thermal tunneling is given by the action \cite{Linde:1981zj}
\begin{align}
\label{eq:bounce_action_S3}
\frac{S_3}{T} = \frac{4\pi}{T}\int dr\, r^2 \left[ \frac{1}{2}\phi^{'}(r)^2 + V(\phi) \right], 
\end{align}
where the tunneling trajectory reads
\begin{equation}
\phi^{''}(r) + \frac{2}{r}\phi^{'}(r) = \frac{dV}{d\phi}, \quad \phi^{'}(0)=0, \quad \underset{r \to \infty}{\textrm{lim}}\,\phi(r) =0. 
\end{equation}
We calculate the bounce action $S_3$ in Eq.~\eqref{eq:bounce_action_S3} both numerically and analytically. 

\begin{figure}[!ht]
\centering
\raisebox{0cm}{\makebox{\includegraphics[width=0.65\textwidth, scale=1]{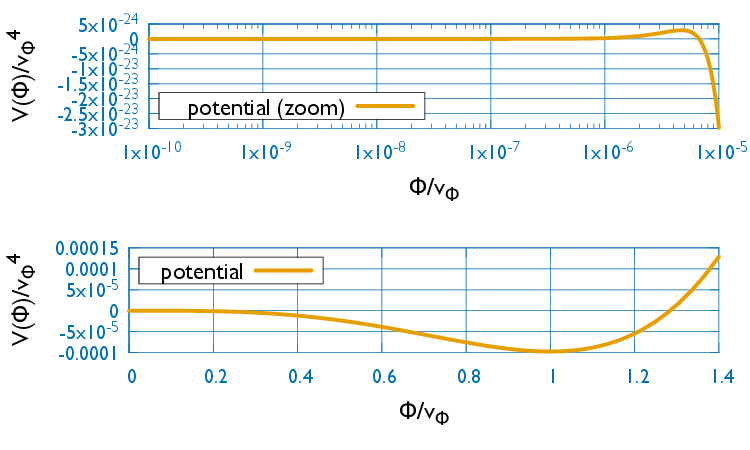}}}
{\makebox{\includegraphics[width=0.49\textwidth, scale=1]{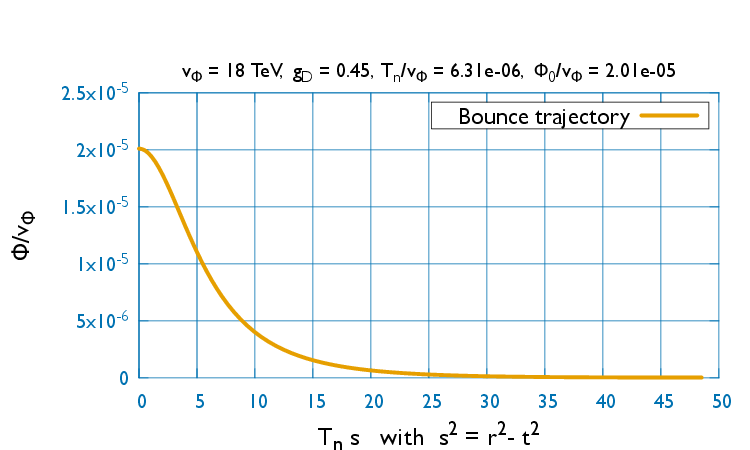}}}
{\makebox{\includegraphics[width=0.49\textwidth, scale=1]{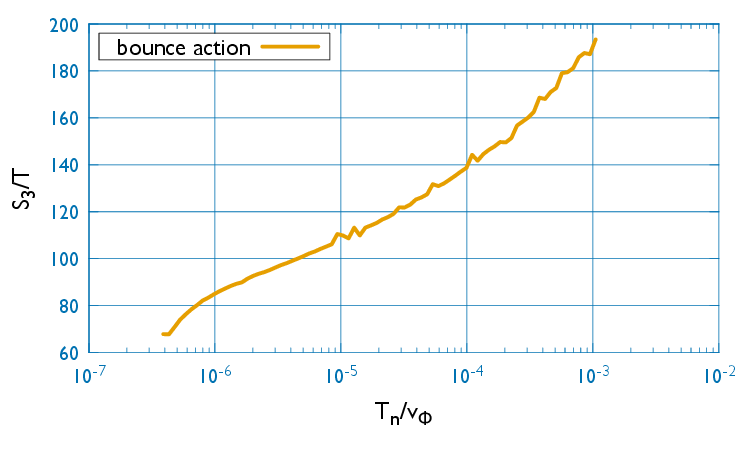}}}
\caption{\normalsize  Outputs of ${\tt CCBounce}$ \cite{cppbounce}: the scalar potential (\textbf{top}), the bounce profile at nucleation (\textbf{bottom left}) and the bounce action for different values of the temperature (\textbf{bottom right}). We have fixed the $U(1)_D$ gauge coupling $g_D = 0.45$ and the $U(1)_D$ scalar vev $v_\phi=17~\rm TeV$. }
\label{fig:CCbounce} 
\end{figure}

\begin{figure*}[ht!]
\centering
\includegraphics[width=250pt]{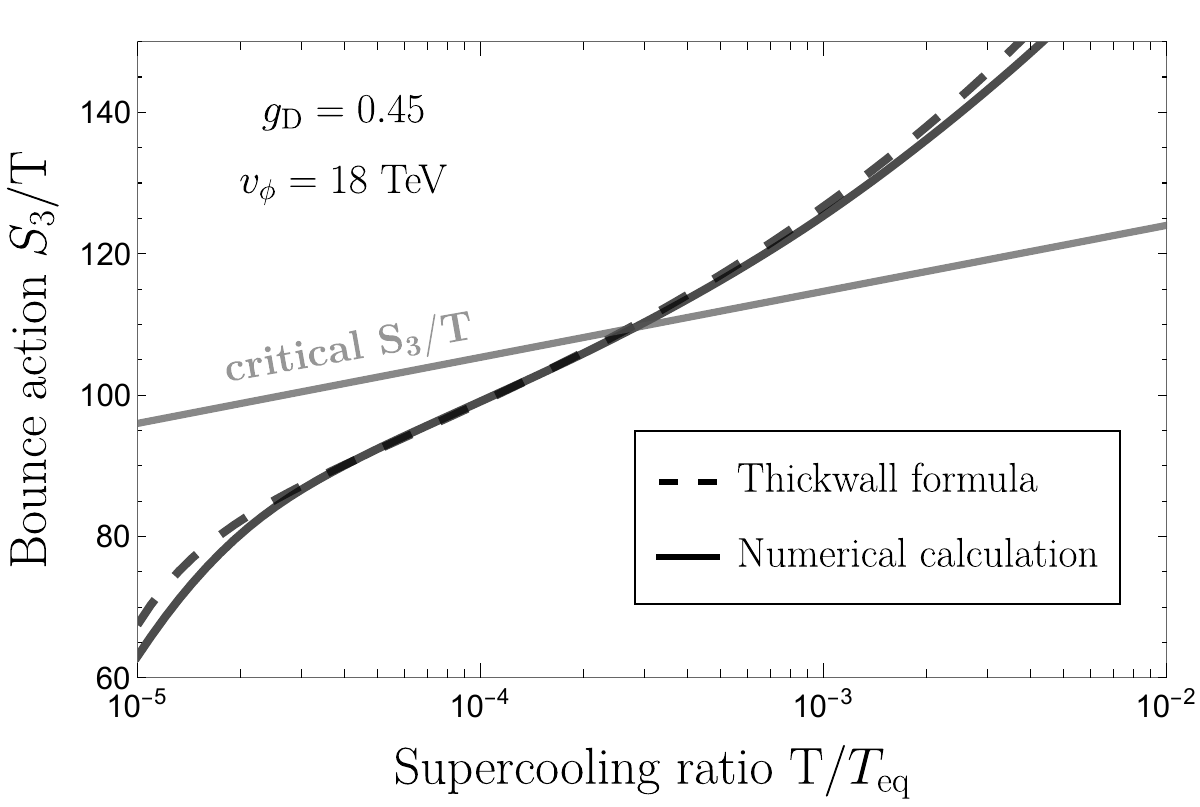}
\includegraphics[width=250pt]{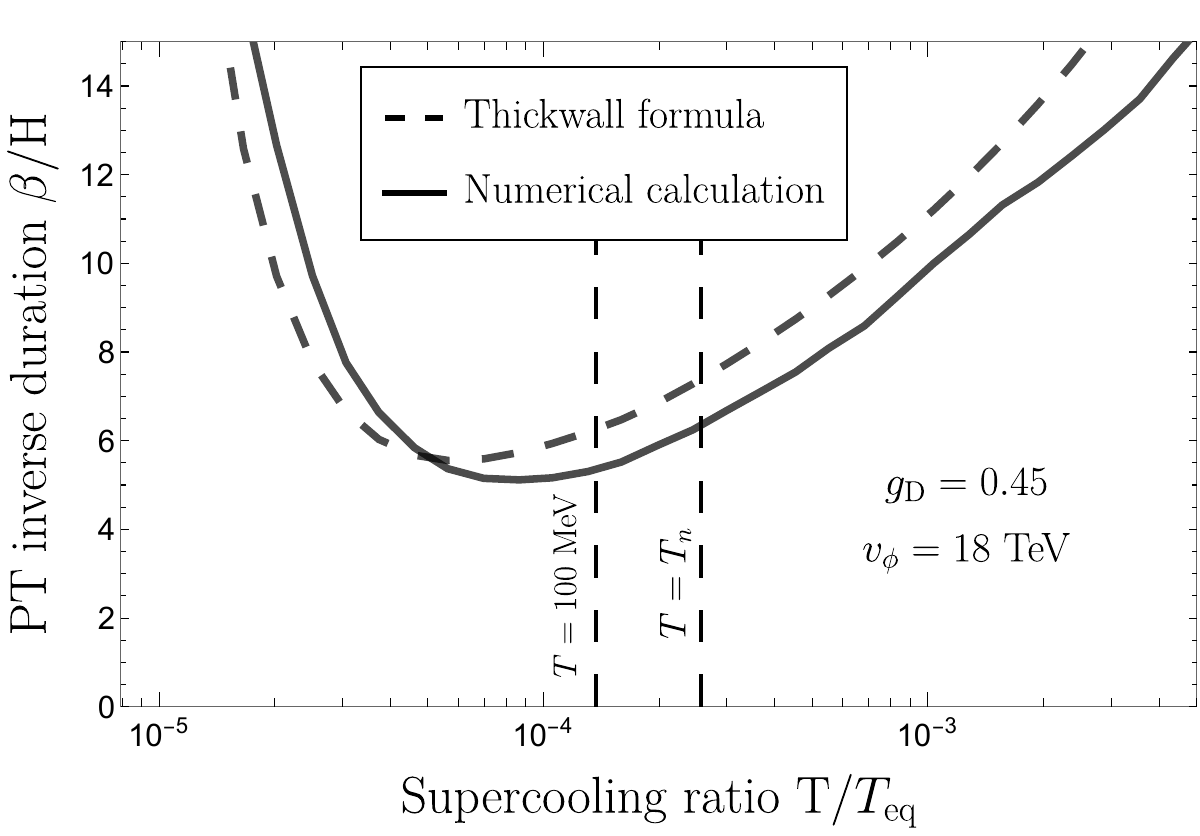}
\caption{\normalsize  Bounce action $S_3/T$ (\textbf{left}) and rate $\beta/H$ of the phase transition (\textbf{right}).  The thickwall analytical formula is a good approximation of the numerical results. }
\label{fig:CW_gX_Tnuc_alpha_beta}
\end{figure*}

\underline{Numerical method} ---
We wrote an over-shooting/under-shooting C++ algorithm {$\tt CCBounce$} \cite{cppbounce} using the full potential in Eq.~\eqref{eq:app_potential_full} and the quartic coupling being given by Eq.~\eqref{eq:lambda_phi}. We checked that the error is below the percent level as explained in \cite[chap.~6]{Gouttenoire:2022gwi}.  Outputs of {$\tt CCBounce$} are displayed in Fig.~\ref{fig:CCbounce}. The bounce action after slight smoothening is shown with thick lines in Fig.~\ref{fig:CW_gX_Tnuc_alpha_beta}.

\begin{figure}[ht!]
\centering
\includegraphics[width=370pt]{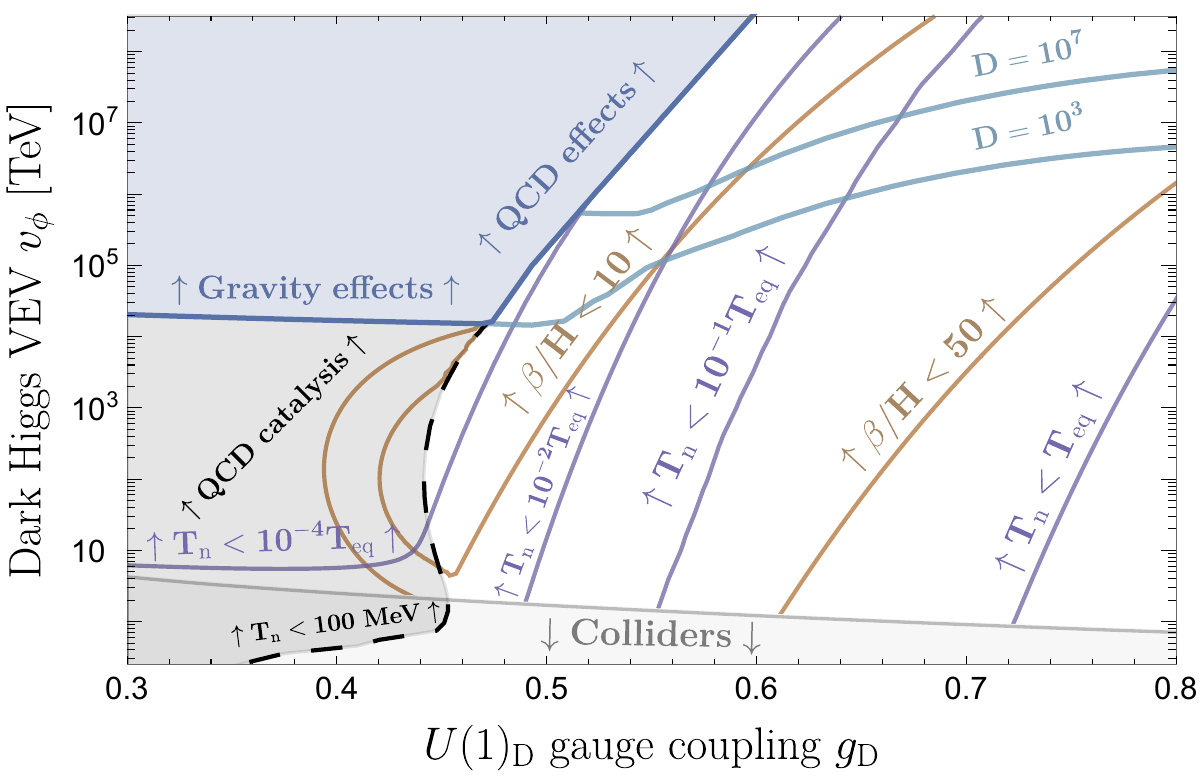}
\caption{\normalsize  Parameter space of the minimal $U(1)_{\rm D}$ scale-invariant model with large amount of supercooling $T_{n}/T_{\rm eq}< 1$ (\textbf{purple} lines) and slow completion rate encoded by $\beta/H$ (\textbf{orange} lines). See Fig.~\ref{fig:PBHDM_vphi_gD_withQCD} for the rest of the legend.}
\label{fig:CL_Tn_beta}
\centering
\includegraphics[width=400pt]{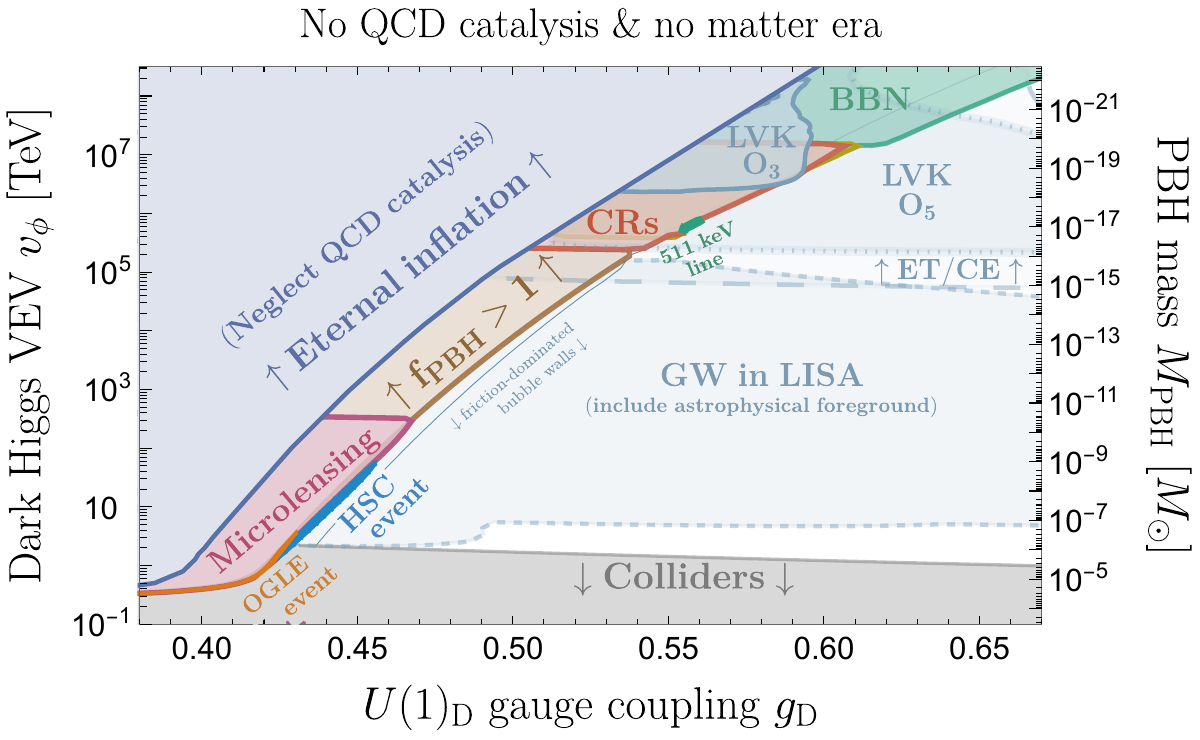}
\caption{\normalsize  Same as Fig.~\ref{fig:PBHDM_vphi_gD_withQCD} in the main text under the assumption that the QCD catalysis does not arise and that the universe follows an adiabatic evolution after the PT. The later point would arise if the dark scalar decays right after percolation, see e.g. \cite{Baldes:2023rqv}. In contrast to the minimal case presented in the main text, all the regions producing observable PBHs are testable with future GW interometers, above astrophysical foregrounds, including the region explaining DM as made of PBHs. The thin orange region in the bottom-left corner can explain the six ultra-short lensing events observed with OGLE telescopes after five years of observation of the Galactic bulge and which could be attributed to PBHs with masses $10^{-6}M_{\odot} > M_{\rm PBH} > 10^{-3}M_{\odot}$ \cite{2017Natur.548..183M,Niikura:2019kqi}. This region is absent in Fig.~\ref{fig:PBHDM_vphi_gD_withQCD}, which implies that effects from QCD catalysis prevent the scale-invariant EWPT to explain  OGLE events. In principle, OGLE events could be explained if QCD corrections to the EW phase transition are weakened for some reasons.}
\label{fig:PBHDM_vphi_gD_noQCD}
\end{figure}

\underline{Thick-wall method} ---
We numerically find that the field value $\phi_*$ at the center of a nucleated bubble is of the order of the temperature $\phi_* \sim T_n \ll v_\phi$. This allows two approximation schemes.
First of all, we can Taylor-expand the potential $V(\phi)=V_{T=0}+V_{\rm T} +V_{{\rm QCD}}$ at first order in $\phi/T \ll 1$:
\begin{equation}
V(\phi) \simeq \frac{m_{\rm eff}^2(T)}{2}\phi^2 - \frac{\lambda_{\rm eff} }{4}\phi^4
\label{eq:effective_pot_cw_HT}
\end{equation}
with 
\begin{equation}
m_{\rm eff}^2(T) \simeq \frac{1}{12}\,g_{\rm D}^2\,T^2 -m_{\mathsmaller{\rm QCD}}^2, \qquad \lambda_{\rm eff}  \simeq \beta_{\lambda}\,\ln{\frac{e^{1/4}\vphi}{\phi}} .
\end{equation}
Second of all, we can use the thick-wall formula \cite{Baldes:2021aph}, see for instance the textbook \cite[chap.~6]{Gouttenoire:2022gwi}:
\begin{equation}
\label{eq:S3_thick}
\frac{S_3}{T} = \frac{16 \pi}{3\,\lambda_{\rm eff} } \frac{m_{\rm eff}}{T}, \qquad \text{and}  \qquad \phi_* = 2\frac{m_{\rm eff}}{\sqrt{-\lambda_{\rm eff}}}.
\end{equation}
In terms of the physical parameters, $S_3$ becomes:
\begin{align}
\label{eq:S3overT_cw_app}
&\frac{S_3}{T} \simeq  \frac{A}{\textrm{log}\left(\frac{M}{T}\right)} \quad \text{with} \quad M = 0.43\,g_{\rm D}\,\vphi, \quad \\
&A= \frac{127.3}{g_{\rm D}^3}\sqrt{1 - \left(\frac{T_{\mathsmaller{\rm QCD}}}{T}\right)^{\!2}}, \quad T_{\mathsmaller{\rm QCD}} =  \frac{\sqrt{6\lambda_{\rm h\sigma}}}{g_{\rm D}}\left<h\right>_{\mathsmaller{\rm QCD}}. \notag
\end{align}
and the field value at the tunneling exit reads:
\begin{equation}
\label{eq:exit_point}
\phi_* \simeq \left(\frac{8}{\log{\frac{v_\phi}{T}}}\right)^{1/2} \frac{T}{g_{\rm D}}.
\end{equation}
\underline{Comparison between $O_3$ and $O_4$-symmetric bounce trajectories} ---
The decay rate of the false vacuum due to $O(3)$- and $O(4)$-symmetric tunneling trajectories is given by \cite{Coleman:1977py, Callan:1977pt,Linde:1980tt, Linde:1981zj}:
\begin{equation}
\Gamma(T) \simeq {\rm max} \left[   \mathcal{A}_3 \,{\rm exp} \left(-S_3/T\right) ,~ \mathcal{A}_4 \,{\rm exp} \left(-S_4 \right)  \right],
\label{eq:tunneling_rate}
\end{equation}
where $\mathcal{A}_3=T^4(S_3/T/2\pi)^{3/2}$ and $\mathcal{A}_4=R_{\rm n}^{-4}(S_4/2\pi)^{2}$ are prefactors and $R_n$ is the bubble radius at nucleation.  As shown in bottom-left panel of Fig.~\ref{fig:CCbounce}, the bubble radius at nucleation is about $R_n\simeq 10/T_n$, hence
\begin{equation}
\textrm{log}\left(\frac{\mathcal{A}_3}{\mathcal{A}_4} \right) \simeq 7.7 + 0.5\,\textrm{log}\left(\frac{120}{S_{\rm crit}} \right),
\end{equation}
where $S_{\rm crit}$ is the critical bounce action (either $S_3/T$ or $S_4$) when nucleation becomes efficient.
The $O_4$-symmetric bounce for quantum tunneling reads in the thick-wall limit \cite{Baldes:2021aph,Gouttenoire:2022gwi}:
\begin{equation}
S_4 \simeq \frac{2 \pi^2}{\lambda_{\rm eff}}.
\label{eq:S4_cw}
\end{equation}
Comparing with Eq.~\eqref{eq:S3_thick}, we conclude that the $O_4$ tunneling rate is always smaller than the $O_3$-symmetric tunneling rate when:
\begin{equation}
\frac{S_3}{T}   \lesssim S_4+\textrm{log}\left(\frac{\mathcal{A}_3}{\mathcal{A}_4} \right) \qquad \rightarrow \qquad m_{\rm eff} \lesssim 1.7 T \qquad \rightarrow \qquad g_{\rm D} \lesssim 5.7, \label{eq:comp_S3_S4_cw}
\end{equation}
Therefore, in the parameter space of interest tunneling is always dominated by the $O_3$-symmetric contribution.  Fig.~\ref{fig:CW_gX_Tnuc_alpha_beta} shows that the thick-wall formula compares reasonably well with the numerical formula. We can safely use it in the rest of this work. The nucleation temperature $T_n$ and phase transition rate $\beta/H$ are plotted in Fig.~\ref{fig:CL_Tn_beta}, and also Fig.~\ref{fig:beta_vs_Tn}.\\

\begin{figure*}[ht!]
\centering
\includegraphics[width=250pt]{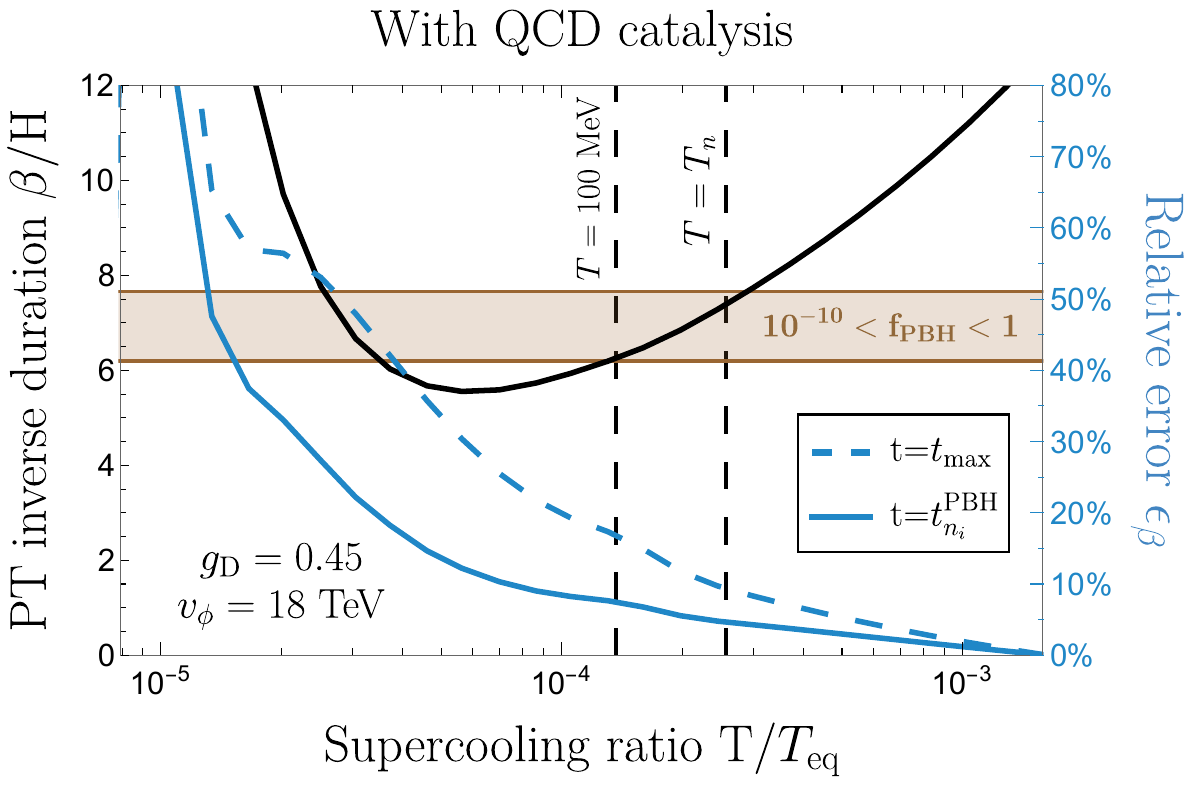}
\includegraphics[width=250pt]{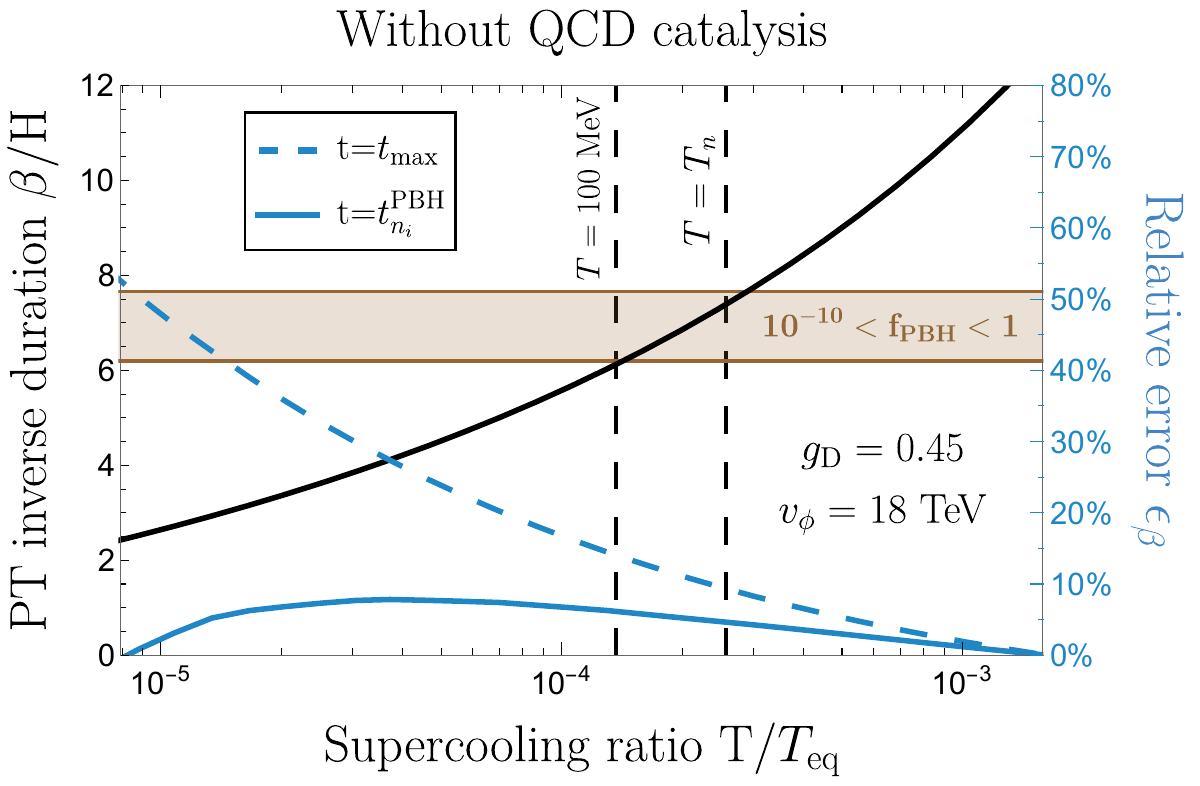}
\caption{\normalsize The blue lines depict the relative error $\epsilon_\beta = \beta_2(t-t_n)/2\beta$ in Eq.~\eqref{eq:2ndorder_betaH_2} resulting from neglecting the second-order correction in the Taylor expansion of the tunneling rate logarithm $\ln \Gamma_{\mathsmaller{\rm V}}(t) \simeq \beta (t-t_n) + \beta_2(t-t_n)^2/2$. This error is evaluated either at the nucleation delay time $t_{n_i}^{\rm PBH}$ or at the time $t_{\rm max}$ when the density contrast $\delta$ reaches its maximum value set to $\delta_c\simeq 0.45$, both of which are determined numerically. The black line represents the values of the PT completion rate $\beta/H$, while the brown region indicates the range that produces observable PBHs. }
\label{fig:relative_error}
\end{figure*}

\underline{Validity of the $\beta$-parametrization of the tunneling rate} ---
We now discuss the validity of truncating the Taylor-expansion of the tunneling rate logarithm at 1st order in $t-t_n$ in Eq.~\eqref{eq:tunneling_rate_def_0}. 
Assuming $\Gamma_{\mathsmaller{\rm V}}(t) \simeq T^4 e^{-S_3/T}$, the Taylor-expansion at 2nd order reads
\begin{equation}
\label{eq:2ndorder_betaH}
\ln\left( \Gamma_{\mathsmaller{\rm V}}(t) \right) \simeq \beta (t-t_n) + \beta_2(t-t_n)^2/2,
\end{equation}
with
\begin{align}
\beta/H = T \frac{d(S_3/T)}{dT},\quad {\rm and} \quad \beta_2/H^2 = -4\frac{\dot{H}}{H^2}- \frac{\beta}{H} + T^2 \frac{d^2(S_3/T)}{dT^2}.
\end{align}
In the large supercooling limit we can set $\dot{H}= 0$, and the ratio of the 2nd order term over the 1st order term becomes
\begin{equation}
\label{eq:2ndorder_betaH_2}
\epsilon_\beta \equiv \frac{\beta_2(t-t_{n})}{2\beta} \simeq \left(\frac{H}{\beta}T^2 \frac{d^2(S_3/T)}{dT^2} -1\right)\frac{H(t-t_{n})}{2}.
\end{equation}
In Fig.~\ref{fig:relative_error}, we show with blue lines the relative error $\epsilon_\beta$ in Eq.~\eqref{eq:2ndorder_betaH_2} evaluated at two characteristic times. First, we consider the time $t_{n_i}^{\rm PBH}$, which corresponds to the time until which nucleation is delayed within the past light cone of the PBH-forming region. Second, we consider the time $t_{\rm max}$, at which the density contrast reaches its maximum value, set equal to the threshold $\delta_c \simeq 0.45$. Both times are determined numerically following App.~\ref{app:PBH_formation}. We conclude that neglecting the second-order correction in the Taylor expansion of the tunneling rate logarithm is a reasonable approximation when nucleation occurs through the Coleman-Weinberg potential alone, without QCD catalysis. Specifically, for values $\beta/H \sim 6$ associated with efficient PBH production, the error $\epsilon_\beta$  is only $5\%$ at $t_{n_i}^{\rm PBH}$ and $15\%$ at $t_{\rm max}$. However, the error $\epsilon_\beta$  can reach $50\%$ at $t_{n_i}^{\rm PBH}$ and $30\%$ at $t_{\rm max}$ when nucleation is catalyzed by QCD confinement. The relative error $\epsilon_\beta$ is expected to increase at smaller values of $v_\phi$, where the effects of QCD are more pronounced. Given the uncertainties in modelling QCD catalysis, we consider this level of precision sufficient for the purposes of this work and leave a more detailed analysis for future studies. 

For future reference, we note that the time $t_{n_i}^{\rm PBH}$ can been estimated analytically from Ref.~\cite[App.~C2]{Gouttenoire:2023naa} as
\begin{equation}
\label{eq:t_ni_PBH}
\beta(t_{n_i}^{\rm PBH} -t_n)\simeq \ln\left(-\frac{3\beta}{4\pi H}\ln\left( \mathcal{P}_{\rm coll} \right)\right),
\end{equation}
where $ \mathcal{P}_{\rm coll}$ is the collapse probability in Eq.~\eqref{eq:proba_coll_ana}. We find that Eq.~\eqref{eq:t_ni_PBH} overestimate its numerically calculated value by less than $50\%$.

\underline{Validity of the single-field approximation} ---
Apart from the QCD contribution, in the above discussion we have neglected the motion of the Higgs field $h$ during tunneling of the $U(1)_{\rm D}$ scalar $\phi$. We now check that this is a valid approximation.
In the supercooling limit $T_n \ll T_{\rm eq}$, the exit point of tunneling $\phi_{*}$ in Eq.~\eqref{eq:exit_point} is hierarchically small:
\begin{equation}
\phi_{*} \simeq \left(\frac{8}{\ln{\frac{v_{\phi}}{T_n}}}\right)^{1/2}\frac{T}{g_{\rm D}} ~\ll~v_{\phi}.
\end{equation}
The motion of $\phi$ during tunneling does induce a motion of $h$ if the negative induced mass $-\frac{1}{2}\lambda_{h\phi}\phi_{*}^2$ is larger than the positive thermal mass $ m_h^2(T) \simeq (\frac{3g_2^2}{16}+\frac{y_t^2}{4}+\frac{\lambda_h}{2})T^2$ which implies:
\begin{equation}
\label{eq:higgs_motion}
g_{\rm D}~<~ 0.4\left(\frac{8}{\ln{\frac{v_{\phi}}{T_n}}}\right)^{1/2} \frac{0.4~\rm TeV}{v_\phi}.
\end{equation}
Whenever Eq.~\eqref{eq:higgs_motion} is verified, during tunneling the Higgs field acquires the value $h_{\rm exit} \simeq \phi_{*}\sqrt{\lambda_{h\phi}/\lambda_h}$.
The motion of $h$ backreacts on the tunneling if the negative induced mass $-\frac{1}{2}\lambda_{h\phi}h_{\rm exit}^2$ is comparable to the thermal mass $m_{\phi}^2(T)\simeq g_{\rm D}^2T^2/12$ which implies:
\begin{equation}
\label{eq:higgs_motion_backreaction}
g_{\rm D}~<~ 0.8\left(\frac{8}{\ln{\frac{v_{\phi}}{T_n}}}\right)^{1/4} \frac{0.4~\rm TeV}{v_\phi},
\end{equation}
where we used $\lambda_{h\phi}=2\lambda_h(v_{\rm EW}/\vphi)^2$ and  $\lambda_h \simeq 0.13$. We conclude that we can safely neglect the motion of the Higgs during tunneling (except for QCD catalysis discussed in the main text).

\subsection{Collider constraints}

\underline{Mixing angle} ---
Within a view to applying collider bounds to the SM extension we consider, we review the derivation of the mixing angle $\theta_{h\phi}$. The Lagrangian after spontaneous symmetry breaking (with quantum corrections included) reads:
	\begin{equation}
	\mathcal{L} \supset -\mu_h^2 |H|^2 -\lambda_h |H|^4- \mu_\phi |\Phi|^2- \lambda_\phi |\Phi|^4- \lambda_{h \phi } |\Phi|^{2} |H|^{2},
	\end{equation}	
with $H = \left(0,\frac{v_{\rm EW}+\tilde{h}}{\sqrt{2}}\right)$ and $\Phi = \frac{v_{\phi}+\tilde{\phi}}{\sqrt{2}}$ the gauge eigenstates. 
	The minimum of the potential at $(v_\phi, \, v_{\rm EW})$ implies:
	\begin{equation}
	\mu_h^2 = -\lambda_h v_{\rm EW}^2 - \frac{1}{2}\lambda_{h\phi} v_{\phi}^2, \quad \mu_\phi^2 = -\lambda_\phi v_{\phi}^2 - \frac{1}{2}\lambda_{h\phi} v_{\rm EW}^2,\qquad v_{\rm EW} \simeq 246~\rm GeV.
	\end{equation}
The mixing angle $\theta_{h\phi}$ is defined as the rotation angle between the gauge eigenstates and the mass eigenstates:
	\begin{equation}
\left( \begin{array}{c}
h \\
\phi
\end{array} \right) = \left( \begin{array}{cc}
\cos{\theta_{h\phi}} & \sin{\theta_{h\phi}}  \\
-\sin{\theta_{h\phi}}  & \cos{\theta_{h\phi} }
\end{array} \right) \left( \begin{array}{c}
\tilde{h} \\
\tilde{\phi}
\end{array} \right),
	\end{equation}
with associated mass eigenvalues:
	\begin{align}
	&m_h^2 = 2 \lambda_h v_{\rm EW}^2 \cos^2{\theta_{h\phi}} + 2 \lambda_\phi v_\phi^2  \sin^2{\theta_{h\phi}}- \lambda_{h\phi} v_\phi v_{\rm EW}  \sin{2\theta_{h\phi}} , \\
	&m_\phi^2 = 2 \lambda_h v_{\rm EW}^2 \sin^2{\theta_{h\phi}} + 2 \lambda_\phi v_\phi^2  \cos^2{\theta_{h\phi}}+ \lambda_{h\phi} v_\phi v_{\rm EW}  \sin{2\theta_{h\phi}} .
	\end{align}
We find:
\begin{align}	
  \label{eq:Higgs_mixing_app}
	\sin{2\theta_{h \phi}} &  = \frac{ \lambda_{ h \phi } v_{\phi} v_{\rm EW} }{ \sqrt{((\lambda_{\phi}v_{\phi}^{2} - \lambda_{h}v_{\rm EW}^{2})^2+(\lambda_{ h \phi } v_{\phi} v_{\rm EW})^2 }} , 
	\end{align}
where $\lambda_{h\phi} = (m_h/v_{\phi})^2$, $\lambda_h = m_h^2/2v_{\rm EW}^2$ and $\lambda_\phi= 11\beta_\lambda/6$.
The dominant dark Higgs decay channel is into two Higgs $\phi \to hh$:
\begin{equation}
\label{eq:Gamma_phi_hh_app}
    \Gamma_{\phi\to hh} = \frac{\lambda_{h\phi}^2 v_\phi^2}{32\pi m_\phi} = 0.3H_n \left( \frac{0.5}{g_{\rm D}}\right)^6 \left( \frac{10^3~\rm TeV}{v_{\phi}}\right)^5,
\end{equation}
where $m_\phi =\beta_\lambda v_\phi$ is the scalar mass and $H_n^2=\Delta V/3M_{\rm pl}^2$ is the Hubble rate at percolation.
The other decay channel, through Higgs mixing, is further suppressed by $(v_{\rm EW}/v_{\phi})^2$:
\begin{equation}
     \Gamma_{\phi\to h\to SM} = \Gamma_h(m_\phi) \sin^2(\theta_{h\phi}) \simeq  \Gamma_h(m_\phi) \left(\frac{3 \lambda_{h\phi} v_{\rm EW}}{11\beta_\lambda v_{\phi}}  \right)^2,
\end{equation}
where $\Gamma_h(m_\phi)$ is the Higgs decay width evaluated at the singlet scalar mass $m_h\to m_\phi$ (we have $\Gamma_h(m_h)  \simeq 4.1~\rm MeV$~\cite{ParticleDataGroup:2020ssz}). \\

\begin{figure}[th!]
\centering
\includegraphics[width=450pt]{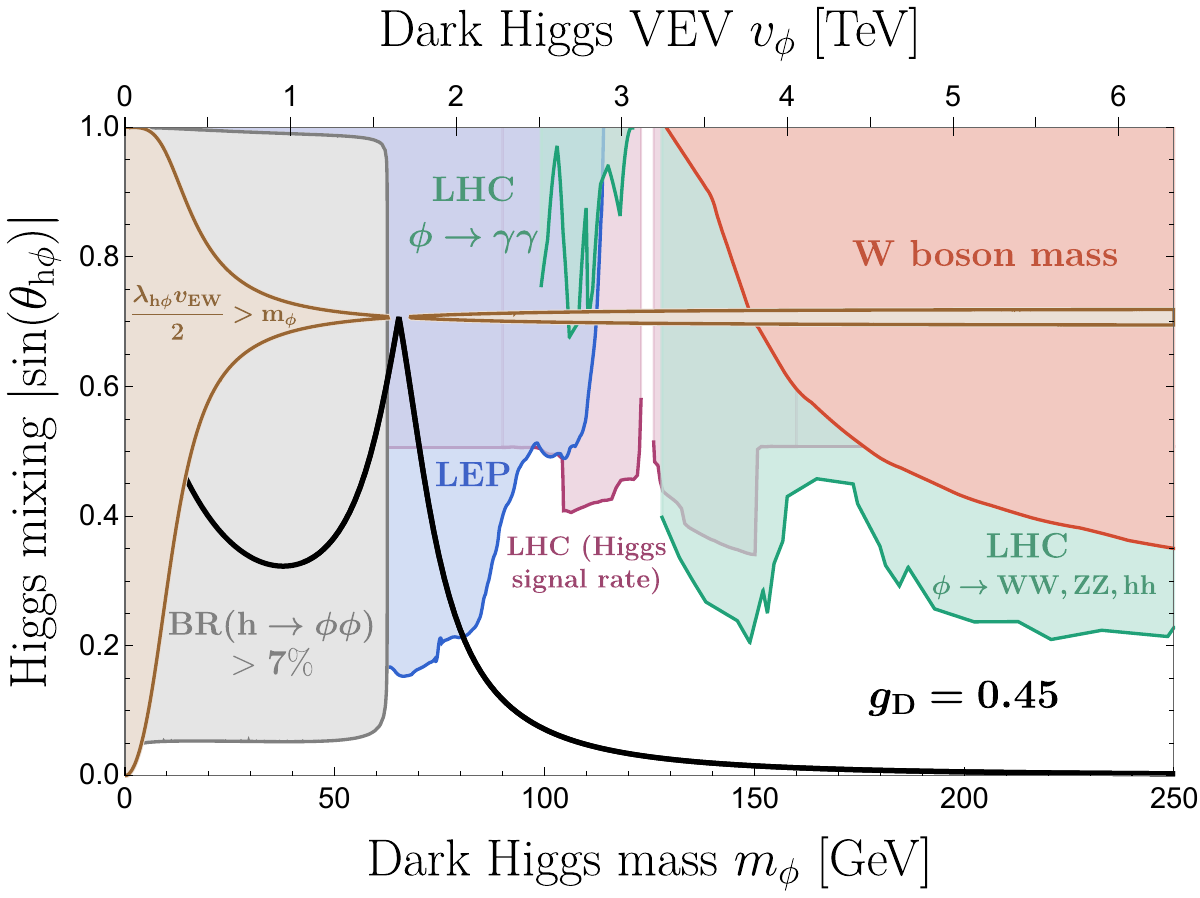}
\caption{\normalsize   \label{fig:collider_constraints} Collider constraints on mixing angle $\theta_{h\phi}$ in scale-invariant $U(1)_{\rm D}$ extension of the SM  (\textbf{black line}) as a function of the extra Higgs mass $m_\phi$. We show LEP searches for 4-fermion final states from Higgs strahlung $e^+e^- \to Z\phi$ \cite{ALEPH:2006tnd} (\textbf{blue}), resonant production at LHC \cite{Ilnicka:2018def} (\textbf{green}), electroweak fit of the W mass \cite{Ilnicka:2018def} (\textbf{red}), Higgs signal shape \cite{Robens:2015gla} (\textbf{purple}), Higgs decay to new particle \cite{Robens:2019kga} (\textbf{gray}). In the \textbf{brown} region, the scalar mass $m_\phi$ is induced by the Higgs VEV.  }
\end{figure}

\underline{Higgs searches} ---
We use Eq.~\eqref{eq:Higgs_mixing_app} to recast existing collider constraints on additional Higgs boson on the parameter space of the minimal scale-invariant $U(1)_{\rm D}$ extension of the SM, see Fig.~\ref{fig:collider_constraints}. For $g_{\rm D} \sim 0.5$, we find that the tightest constraints are given by LEP searches $v_{\phi} \gtrsim 1.5 ~\rm TeV$, the mass of the dark Higgs being $m_\phi \gtrsim 80 ~\rm GeV$.

\section{Bubble dynamics}
\label{app:GW}

\subsection{Energy budget}
The motion of bubble walls is subject to the driving vacuum pressure $\Delta V$ and to the retarding friction pressure {induced by the bremsstrahlung emission of vector bosons $V$ by scalar particles $\phi$ entering bubbles. After resumming leading-logs quantum corrections (LL), the friction pressure reads }
\cite{Bodeker:2017cim,Azatov:2020ufh,Gouttenoire:2021kjv,Azatov:2023xem}:
\begin{equation}
\label{eq:PLL}
\mathcal{P}_{\rm LL} = c_0\,g_{\rm D}^3 \gamma v_{\phi} T_{n}^3 \log\left( \frac{v_{\phi}}{T_n}\right),\qquad c_0=\mathcal{O}(1),
\end{equation}
where $\gamma$ is the bubble wall Lorentz factor. We suppose the leading-order pressure $\mathcal{P}_{\rm LO}$ to be sub-dominant since it does not grow with $\gamma$ \cite{Bodeker:2009qy}.
In the absence of friction $\mathcal{P}_{\rm LL} \ll \Delta V$, all the vacuum energy $E_{\rm vac}=4\pi R^3\Delta V/3$ of a bubble of radius $R$ is converted into kinetic energy of the wall $E_{\rm wall} = 4\pi R^2\sigma \gamma$, where $\sigma$ is the wall surface tension. In that case, the wall Lorentz factor at collision reaches the maximal value allowed by energy conservation:
\begin{equation}
\label{eq:gamma_run}
    \gamma_{\rm max} = \frac{\Delta V R_{\rm coll}}{3\sigma} = \frac{2R_{\rm coll}}{3R_{\rm crit}}.
\end{equation}
The mean bubble radius at collision is given by $R_{\rm coll} \simeq \pi^{1/3}/\beta$ \cite{Enqvist:1991xw}. We introduced $R_{\rm crit}=2\sigma/\Delta V$ which is the critical radius for nucleating a bubble, found as the saddle point of the free energy $F(R)=4\pi \sigma R^2-4\pi R^3\Delta V/3$ in the  thin-wall approximation.
As bubble walls accelerate, the retarding pressure $\mathcal{P}_{\rm LL}$ in Eq.~\eqref{eq:PLL} grows linearly with $\gamma$. The walls stops accelerating as soon as $\mathcal{P}_{\rm LL} \simeq \Delta V$, with the associated Lorentz factor:
\begin{equation}
\label{eq:gamma_LL}
    \gamma_{\rm LL} \simeq \frac{\Delta V}{c_0\,g_{\rm D}^3 \gamma v_{\phi} T_{n}^3 \log\left( \frac{v_{\phi}}{T_n}\right)}.
\end{equation}
Whether the latent heat of the PT is dominantly converted into the plasma or into the kinetic energy of the wall depends on whether Eq.~\eqref{eq:gamma_run} or \eqref{eq:gamma_LL} dominate. The Lorentz factor at bubble collision is given by the formula:
\begin{equation}
    \gamma_{\rm coll} = \textrm{Min}\left[\gamma_{\rm max},~ \gamma_{\rm LL}\right].
\end{equation}
The fraction of the latent heat converted into wall kinetic energy reads (see also \cite{Ellis:2019oqb}):
\begin{equation}
\label{eq:kappa_coll_app}
    \kappa_{\rm coll} = \frac{E_{\rm wall}}{E_{\rm vac}} = \frac{3\gamma \sigma R_{\rm coll}}{\Delta V} = \frac{\gamma_{\rm coll}}{\gamma_{\rm max}}.
\end{equation}

\subsection{Reheating after a matter era}
\label{app:reheating_matter}
At the time of bubble percolation, the latent heat of the universe has been converted into two fluids, the radiation-like energy density $\rho_{\rm shock}$ of the ultra-relativistic shock waves generated by the work of the friction pressure, and the radiation-like energy density of the scalar field gradient $\rho_{\rm \phi}$. Both are highly peaked distribution of energy-momentum tensor. After that walls pass through each other, the highly peaked energy stored in the wall $E_{\rm wall} \simeq \gamma_{\rm coll}\sigma R_{\rm coll}^2$ is converted into a broadly-distributed oscillating scalar field condensate with energy $E_{\rm osc} \simeq \Delta V R_{\rm coll}^2 \Delta R$ where $\Delta R$ is the distance between the two peaks in energy-momentum tensor after their collision. The radiation-like energy stored in the collided walls is converted into the matter-like oscillating scalar condensate $E_{\rm wall} = E_{\rm osc} $ after the time:
\begin{equation}
\label{eq:Delta_R_max}
    \Delta R_{\rm max} \simeq \frac{\sigma \gamma_{\rm coll}}{\Delta V} \simeq \frac{\gamma_{\rm coll}}{\gamma_{\rm max}}R_{\rm coll} \simeq \kappa_{\rm coll} R_{\rm coll}.
\end{equation}
If the scalar field is long-lived $\Gamma_{\phi}\ll H$, then the scalar field energy density starts redshifting like matter with an energy density after percolation given by:
\begin{equation}
\label{eq:rho_M}
    \rho_{\rm M}(t) = \kappa_{\rm coll} \Delta V \left( \frac{a(t_n)}{a(t)}\right)^3.
\end{equation}
The radiation energy density reads:
\begin{equation}
\label{eq:rho_R}
    \rho_{\rm R}(t)=(1-\kappa_{\rm coll})\Delta V \left( \frac{a(t_n)}{a(t)}\right)^4.
\end{equation}
The scalar field decays after the time approximately given by its inverse decay width:
\begin{equation}
    t_{\rm dec} \simeq \Gamma_{\phi}^{-1},\quad \implies \quad T_{\rm dec} = \frac{1.2}{g_*^{1/4}}\sqrt{M_{\rm pl} \Gamma_{\phi}},
\end{equation}
where we also gave the associated temperature.
If long-lived enough, this would lead to a matter-domination era starting at the time and temperature:
\begin{equation}
\label{eq:T_dom}
    t_{\rm dom} \simeq t_n/\kappa_{\rm coll}^2\quad \implies \quad T_{\rm dom} \simeq \kappa_{\rm coll} T_{\rm eq},
\end{equation}
obtained from equating Eqs.~\eqref{eq:rho_M} and \eqref{eq:rho_R} with $a\propto t^{1/2}$. We conclude that if $t_{\rm dom}<t_{\rm dec}$ then the PT is followed by a matter era starting at $t_{\rm dom}$ and ending at $t_{\rm dec}$.
A transition from a matter to radiation inject an amount of entropy in the universe, e.g. \cite{McDonald:1989jd,Cirelli:2018iax,Gouttenoire:2023roe}, whose magnitude is given by the dilution factor: 
\begin{equation}
\label{eq:dilution_fac_app}
    D \equiv \frac{S_f}{S_i}= 1+ \frac{T_{\rm dom}}{T_{\rm dec}},
\end{equation}
where $S_i$ and $S_f$ is the total entropy in the universe just before and just after the decay of the scalar field, assuming an instantaneous decay. The second equality comes from evolving matter and radiation from $T_{\rm dom}$ to $T_{\rm dec}$. The entropy injection dilutes the abundance of any relic present in the universe before $t_{\rm dom}$. This is the case of PBHs and GWs. For PBHs, the abundance is simply rescaled by:
\begin{equation}
    f_{\rm PBH}\rightarrow \frac{f_{\rm PBH}}{D}.
\end{equation}
For GWs, the spectrum is rescaled by:
\begin{equation}
    \Omega_{\rm GW}(f) \rightarrow \frac{\Omega_{\rm GW}(D^{1/3} f)}{D^{4/3}}S_M(f).
\end{equation}
where the $D$ factors trivially correct for the redshift of the peak amplitude and frequency \cite{Ertas:2021xeh}. Instead, the effects behind $S_M(f)$ are more subtle and consist to change the spectral slope to $\propto f^1$ \cite{Barenboim:2016mjm,Domenech:2020kqm,Ellis:2020nnr,Hook:2020phx} instead of $f^3$ \cite{Durrer:2003ja,Caprini:2009fx,Cai:2019cdl} for modes with wavenumber $k=2\pi f$ which are super-horizon $k < H_n/a$ at percolation and enter the causal horizon  $k = H/a$ during the matter era. In other words, $S_M(f)$ is designed to change the spectral slope to:
\begin{equation}
\label{eq:S_M}
    \Omega_{\rm GW}(f)S_M(f)\propto f^1 \qquad \textrm{for}\qquad f_{\rm dec}< f<f_{\rm dom},
\end{equation}
 assuming continuity with the standard spectrum at $f_{\rm dom}$. We have introduced the frequencies corresponding to the causal horizon $H/2\pi$ at the beginning and end of the matter era red-shifted up to today:
\begin{align}
\label{eq:f_dom_f_dec}
& f_{\rm dom} =\left( \frac{a_{\rm dom}}{a_0}\right) \frac{H_{\rm dom}}{2\pi}, \\
 &f_{\rm dec} =\left( \frac{a_{\rm dec}}{a_0}\right) \frac{H_{\rm dec}}{2\pi}, 
\end{align}
with $H\simeq 0.3g_*^{1/2}T^2/M_{\rm pl}$ given by Friedman's equation and:
\begin{align}
\label{eq:redshift_fac_f_dom_f_dec}
&\frac{a_{\rm dom}}{a_0} =  \frac{1.65 \times 10^{-2}~{\rm mHz}}{D^{1/3}}~\left(\frac{T_{\rm dom}}{100~\rm GeV}\right) \left( \frac{g_{*}(T_{\rm dom})}{100} \right)^{1/6} H_{\rm dom}^{-1},\\
&\frac{a_{\rm dec}}{a_0} =  1.65 \times 10^{-2}~{\rm mHz}~\left(\frac{T_{\rm dec}}{100~\rm GeV}\right) \left( \frac{g_{*}(T_{\rm dec})}{100} \right)^{1/6} H_{\rm dec}^{-1},
\end{align}
where the dilution factor $D$ is only applied to $f_{\rm dom}$.
The GW spectrum  $\Omega_{\rm GW}(f)$  from bubble dynamics is presented in the next section.

\begin{figure}[ht!]
\centering
\includegraphics[width=500pt]{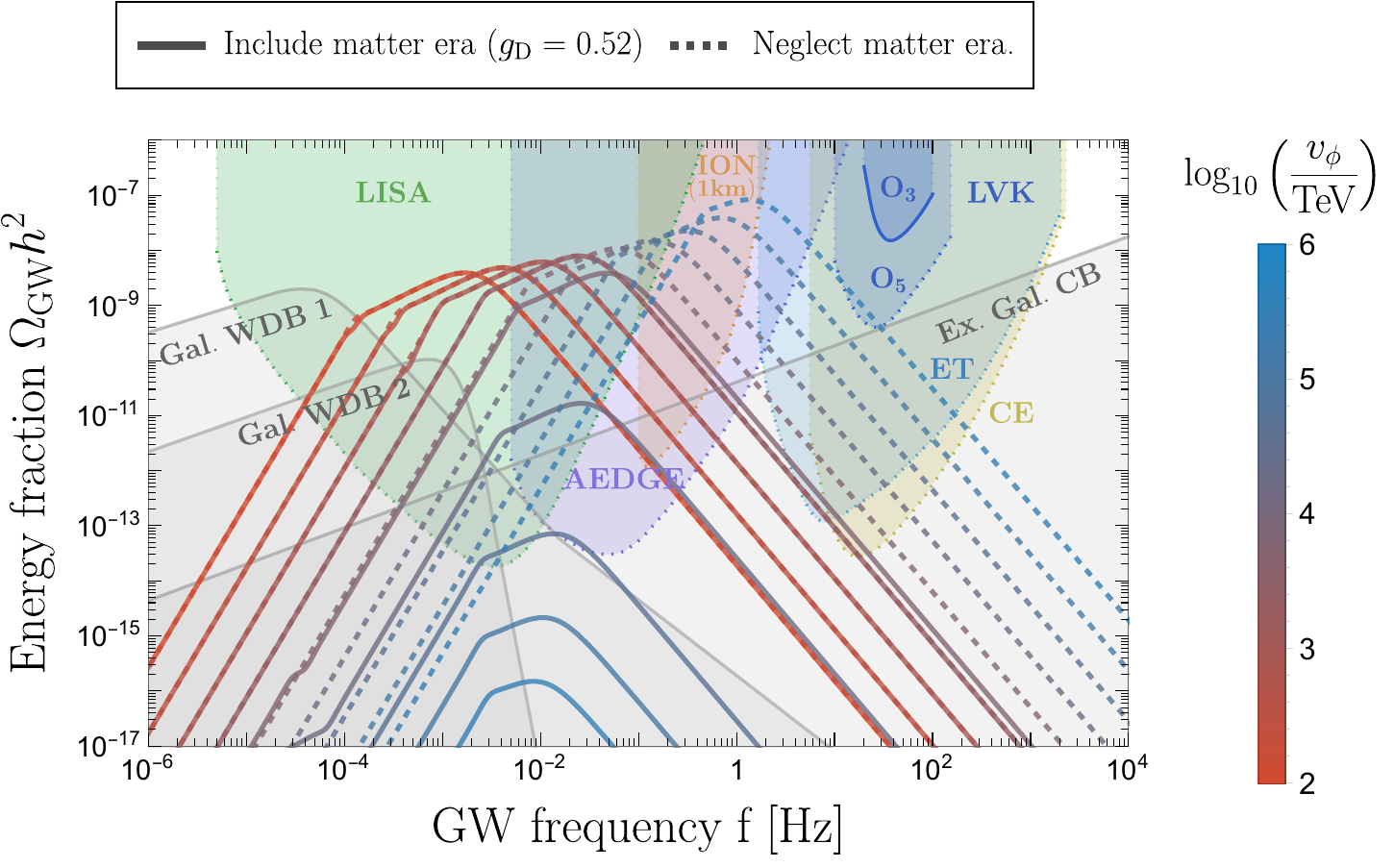}
\caption{\normalsize  Gravitational waves spectrum produced by the supercooled first-order phase transition predicted in the minimal $U(1)_{\rm D}$ extension of the SM. \textbf{Solid} lines shows the GW spectra accounting for the matter-dominated period after the PT, which suppresses the signal at large dark Higgs VEV $v_{\phi}\gtrsim 10^4~\rm TeV$. In comparison, \textbf{dotted} lines show the same GW spectra without the matter era. The \textbf{colorful} regions show the existing constraints from Earth-based laser interometers LIGO-Virgo-Kagra (LVK) run O3 \cite{KAGRA:2021kbb}, assuming Signal-to-Noise Ratio (SNR) detection thresholds of 2, and prospective constraints from LIGO run $O_5$ \cite{LIGOScientific:2014pky}, their follow-ups Einstein Telescope (ET) \cite{Punturo:2010zz,Maggiore:2019uih} and Cosmic Explorer (CE) \cite{Reitze:2019iox}, space-based laser interferometer LISA \cite{Audley:2017drz,Robson:2018ifk,LISACosmologyWorkingGroup:2022jok}, Earth-based atom interferometer AION \cite{Badurina:2019hst,Proceedings:2023mkp} in its space version AEDGE \cite{AEDGE:2019nxb}. 
xxFor LISA, ET and CE, we optimistically assume a low SNR threshold of 1 (equivalent to a 1$\sigma$ deviation from noise) after 20 years of observation \cite{Schmitz:2020syl}. The \textbf{gray} regions show the expected astrophysical foreground from galactic white dwarf binaries (WD B.) according to model 1 \cite{Lamberts:2019nyk,Boileau:2021gbr} and 2 \cite{Robson:2018ifk}, and extragalactic compact (neutron stars and black holes) binaries (Compact B.) fitted on LIGO O3 data \cite{KAGRA:2021kbb}. Instead, the contribution from extragalactic supermassive black holes binaries lies at lower frequencies \cite{Rosado:2011kv}. All the GW constraints in Fig.~\ref{fig:PBHDM_vphi_gD_noQCD} and Figs of the main text are derived assuming that the GW spectra is above the astrophysical foregrounds inside the frequency window of experiments.}
\label{fig:GW_spectra_sens}
\end{figure}

\subsection{GW signal}

\underline{Scalar field gradient} ---
A fraction $\kappa_{\rm coll}$ given  by Eq.~\eqref{eq:kappa_coll_app} of the latent heat is stored in the scalar field gradient localised in bubble walls.
Initially, the GW spectrum has been calculated in the ``envelop'' approximation where walls are infinitely thin and collided parts are removed from the calculation \cite{Kamionkowski:1993fg,Caprini:2007xq, Huber:2008hg,Jinno:2016vai,Weir:2016tov}. The collided part were studied later both analytically \cite{Jinno:2017fby} and numerically \cite{Konstandin:2017sat,Lewicki:2020jiv,Lewicki:2020azd,Cutting:2020nla} in what is known as the ``bulk flow'' model. It is found that the collided part continue to source GW after collision, generating the slow decreasing IR slope $\Omega_{\rm GW}\propto f^{1}$.
The GW spectrum in the bulk flow model gives ($v_w=1$)~\cite{Konstandin:2017sat}:
	\begin{equation}
 \label{eq:Bulk_flow}
	 \Omega_{\rm GW}h^2 \simeq  \frac{1}{D^{4/3}}\frac{10^{-6}}{(g_*/100)^{1/3}} \left(\frac{H_n}{\beta} \right)^{\!2} \left( \frac{\alpha}{1+\alpha} \right)^{\!2}  S(f)S_{H}(f)S_{M}(f),
	\end{equation} 
with the redshift factor between percolation ``$n$'' and today ``$0$'':
	\begin{equation}
	\label{eq:redshift_fac}
a_n/a_0 =  \frac{1.65 \times 10^{-2}~{\rm mHz}}{D^{1/3}}~\left(\frac{T_{\rm eq}}{100~\rm GeV}\right) \left( \frac{g_{*}(T_{\rm eq})}{100} \right)^{1/6} H_{n}^{-1},
	\end{equation}
and the spectral shape $S(f)$ peaking on $f_p$:
	\begin{equation}
 \label{eq:spectral_shape_scalar}
	S(f) = \frac{ 3(f/f_{\rm p})^{0.9} }{2.1+0.9(f/f_{\rm p})^{3}},\quad f_{\rm p} = \left(\frac{a_n}{a_0}\right) 0.8 \left(\frac{\beta}{2\pi}\right).
	\end{equation}
 The dilution factor $D\geq 1$ given by Eq.~\eqref{eq:dilution_fac_app} accounts for the additional redshift due to entropy injection in presence of an eventual early matter era if the scalar $\phi$ is long-lived.
We added the correction factor:
\begin{equation}
\label{eq:Hubble_expansion_fac}
S_{H}(f) = \frac{(f/f_H)^{2.1}}{1+(f/f_{H})^{2.1}}, \quad f_{H} = c_*\left(\frac{a_n}{a_0}\right)\left(\frac{H_n}{2\pi}\right),
\end{equation}
with $c_* = \mathcal{O}(1)$ to impose the scaling $\Omega_{\rm GW}\propto f^3$ for emitted frequencies smaller than the Hubble factor $ H_{\ast}/(2\pi)$ and entering during radiation~\cite{Durrer:2003ja,Caprini:2009fx,Cai:2019cdl,Hook:2020phx}. We set $c_*=1$ and defer the determination of $c_*$ for future studies, see e.g. \cite{Zhong:2021hgo,Giombi:2023jqq}. The factor $S_{\rm M}(f)$ corrects the causality tail $f^3\to f^1$ for modes entering during the matter era instead of the radiation era, as explained in Eq.~\eqref{eq:S_M}.

\underline{Ultrarelativistic shells} ---
We know discuss GW generation from the fraction $1-\kappa_{\rm coll}$ of latent heat stored in terms of fluid motions. In the ultra-relativistic limit, one expects bubble walls to be followed by ultra relativistic shells of particles which are generated by plasma/wall interactions \cite{Jinno:2019jhi,Baldes:2020kam,Azatov:2020ufh,Gouttenoire:2021kjv,Jinno:2022fom,Baldes:2023fsp,Baldes:2023fsp,Azatov:2023xem} which are extremely thin \cite{Baldes:2020kam} and whose impact on the GW spectrum is not yet understood \cite{Cutting:2019zws,Lewicki:2022pdb} at the time of writing. Waiting for future studies to investigate the precise GW spectrum from ultrarelativistic shells, we posit that from the point of view of gravity, the GW production from extremely thin shells of scalar field gradient should be indistinguishable from extremely thin shells of ultra-relativistic particles.\footnote{The author thanks Ryusuke Jinno for useful discussion regarding this point.}
A difference could arise however due to ultra-relativistic shocks being potentially long-lived \cite{Jinno:2019jhi} while scalar field gradient having a reduced lifetime given by Eq.~\eqref{eq:Delta_R_max}. One expect such difference to enhance the GW spectrum by a factor $\beta/H$. Since this is a small factor in the regime of interest for PBH production, we do not include it in our plot. However, we checked that the conclusion of this paper do not depend on this detail. Additional enhancement due to turbulence \cite{Gogoberidze:2007an,Caprini:2009yp,RoperPol:2019wvy,Niksa:2018ofa,Auclair:2022jod} and second order GWs \cite{Domenech:2019quo,Domenech:2021ztg} are foreseeable and left for future works.

Hence, we suppose that the bulk flow model holds in all the parameter space. However, the friction still has an important impact on the GW spectrum through the dependence of the dilution factor on the latent heat fraction into scalar field gradient, see Eqs.~\eqref{eq:dilution_fac_app}, \eqref{eq:T_dom} and \eqref{eq:kappa_coll_app}. We show a family of GW spectrum for different values of PT scale $v_{\phi}$  in Fig.~\ref{fig:GW_spectra_sens} together with LISA reach and astrophysical foregrounds. The constraints on the scale invariant $U(1)_{\rm D}$ extension of the SM are shown in Figs.~\ref{fig:PBHDM_vphi_gD_withQCD} and \ref{fig:PBHDM_vphi_gD_noQCD}. For large dark Higgs VEV $v_\phi \gtrsim 10^4~\rm TeV$, the extended dark Higgs lifetime generates entropy injection and suppresses all GW signals, leaving PBHs as the only possible signatures.

\clearpage
\newpage

\bibliography{biblio}

\end{document}